\documentclass[twocolumn]{aastex631}
\usepackage{xcolor}

\usepackage{CJKutf8}
\usepackage[utf8]{inputenc} 
\usepackage[T1]{fontenc}
\usepackage{ae,aecompl}
\usepackage{graphicx}
\usepackage{amsmath,amsfonts,amssymb,bm}
\usepackage{mathtools}
\usepackage{enumitem}

\usepackage{orcidlink}

\usepackage{txfonts}
\usepackage{color, soul}
\sethlcolor{yellow}
\soulregister\cite7
\soulregister\citep7
\soulregister\ref7
\soulregister\eqref7
\newcommand{\revtext}[1]{#1} 
\newcommand{\reveq}[1]{#1}

\newcommand{\revtexttwo}[1]{#1} 

\DeclareMathOperator{\atantwo}{atan2}

\usepackage{pgfplots}
\usepgfplotslibrary{patchplots}
\pgfplotsset{compat=1.15}

\begin{document}
\begin{CJK*}{UTF8}{bkai}
\title{A hybrid scheme for fuzzy dark matter simulations combining the Schrödinger and Hamilton-Jacobi-Madelung equations}

\author{Alexander Kunkel\orcidlink{0009-0007-2957-4277}}
\email{kunkel.alexander@protonmail.com}
\affiliation{Institute of Astrophysics, National Taiwan University, Taipei 10617, Taiwan}
\affiliation{Department of Physics, National Taiwan University, Taipei 10617, Taiwan}

\author{Hei Yin Jowett Chan\orcidlink{0000-0001-9053-6922}}
\affiliation{Physics Division, National Center for Theoretical Sciences, Taipei 10617, Taiwan}

\author{Hsi-Yu Schive (薛熙于)\orcidlink{0000-0002-1249-279X}}
\email{hyschive@phys.ntu.edu.tw}
\affiliation{Institute of Astrophysics, National Taiwan University, Taipei 10617, Taiwan}
\affiliation{Department of Physics, National Taiwan University, Taipei 10617, Taiwan}
\affiliation{Center for Theoretical Physics, National Taiwan University, Taipei 10617, Taiwan}
\affiliation{Physics Division, National Center for Theoretical Sciences, Taipei 10617, Taiwan}

\author{Hsinhao Huang (黃新豪)\orcidlink{0000-0002-7368-1324}}
\affiliation{Institute of Astrophysics, National Taiwan University, Taipei 10617, Taiwan}
\affiliation{Department of Physics, National Taiwan University, Taipei 10617, Taiwan}

\author{Pin-Yu Liao (廖品瑜)\orcidlink{0009-0007-8469-5880}}
\affiliation{Institute of Astrophysics, National Taiwan University, Taipei 10617, Taiwan}
\affiliation{Department of Physics, National Taiwan University, Taipei 10617, Taiwan}



\begin{abstract}
This paper introduces a hybrid numerical scheme for the fuzzy dark matter model: It combines a wave-based approach to solve the Schrödinger equation using Fourier continuations with Gram polynomials and a fluid-based approach to solve the Hamilton-Jacobi-Madelung equations. This hybrid scheme facilitates zoom-in simulations for cosmological volumes beyond the capabilities of wave-based solvers alone and accurately simulates the full nonlinear dynamics of fuzzy dark matter. We detail the implementation of a Hamilton-Jacobi-Madelung solver, the methodology for phase matching at fluid-wave boundaries, the development of a local pseudospectral wave solver based on Fourier continuations, new grid refinement criteria for both fluid and wave solvers, an interpolation algorithm based on Fourier continuations, and the integration of these building blocks into the adaptive mesh refinement code \texttt{GAMER}. The superiority of the scheme is demonstrated through various performance and accuracy tests, tracking the linear power spectrum evolution in a $10$ Mpc/h box, and a hybrid cosmological simulation in a $5.6$ Mpc/h box. The corresponding code is published as part of the \texttt{GAMER} project on \url{https://github.com/gamer-project/gamer}.
\end{abstract}


\section{Introduction} \label{sec:intro}

\end{CJK*}

The fuzzy dark matter (FDM) model \citep{Hu2000} describes dark matter as a bosonic scalar field composed of ultralight particles with masses typically around $m \sim 10^{-22}$--$10^{-20}$ eV. These particles have negligible self-interactions and possess a macroscopic de Broglie wavelength. In the non-relativistic regime, the dynamics of FDM is governed by the Schrödinger-Poisson equations. Due to its ultralight mass, FDM has an astrophysically relevant de Broglie wavelength on the order of tens of parsecs to a few kiloparsecs. FDM represents a small-scale modification of cold dark matter (CDM), governed by the particle mass. Its wave-like behavior suppresses structure formation on small scales, while CDM-like behavior re-emerges on larger scales. Consequently, the FDM model holds the potential to address the small-scale challenges of CDM \citep{Weinberg2013, Bullock2017}.

Advancements in the understanding of FDM dynamics have been significantly driven by numerical simulations. Various numerical studies have explored structure formation in the FDM model, employing both the wave and fluid formulations. These studies utilize a variety of algorithms such as spectral methods \citep{Mocz2017, Edwards2018, Du2018}, finite difference methods \citep{Schive2014, Schwabe2020, Mina2020}, finite volume methods \citep{Li2018, Hopkins2018}, and smoothed particle hydrodynamics algorithms \citep{Mocz2015, Nori2018}.

\textcite{Woo2009} pioneered the use of high-resolution cosmological simulations of the wave formulation of FDM, utilizing a spectral method with a uniform mesh resolution of $1024^3$ grid points. However, this resolution proved insufficient for probing the innermost regions of halos. The vast dynamical range required for cosmological simulations, spanning from a few megaparsecs down to scales of approximately ten parsecs, demands substantial computational effort. \textcite{Schive2014} addressed this challenge with the code \texttt{GAMER} \citep{Schive2010, Schive2018}. \texttt{GAMER} utilizes an adaptive mesh refinement (AMR) framework to solve the wave formulation of the Schrödinger-Poisson equations. The version of \texttt{GAMER} used in this work, v2.2.1, is available in Zenodo \citep{10.5281/zenodo.15036957}.

A primary limitation in wave-based simulations of FDM arises from the necessity to resolve the de Broglie wavelength, even in regions where the density field is relatively smooth and non-vanishing. This requirement stems from the relationship between the velocity field and the phase gradient of the wave function. Neglecting to resolve the de Broglie wavelength adequately can lead to inaccuracies in the velocity field \citep{Li2018}. Unlike conventional CDM simulations, where higher spatial resolutions are primarily required in denser regions, FDM simulations demand fine resolution across a broader range of scales. In practical terms, for an FDM particle with $m = 10^{-22}$ eV and a velocity of $v = 100$ km/s, the de Broglie wavelength ($\lambda_{dB} \sim 1.2$ kpc) is significantly smaller than the box size required for a cosmological simulation of relevant size. Consequently, wave simulations have traditionally been constrained to relatively small box sizes of a few megaparsecs. However, \cite{May2021, May2022} have started to push these boundaries, achieving a box size of $10$ Mpc/h in the wave formulation for $m = 7\times 10^{-23}$ eV. Still, their simulations had to be stopped at redshift $z=3$ because of insufficient resolution. 

In contrast, simulations utilizing the fluid formulation of FDM are not bound by the requirement to resolve the de Broglie wavelength for capturing large-scale dynamics. Moreover, this advantage allows these simulations to adopt a Lagrangian perspective, incorporating the quantum pressure through smoothed particle hydrodynamics methods. Consequently, the attainable simulation volumes in fluid formulations of FDM are more comparable to those reached in traditional N-body and smoothed particle hydrodynamics approaches used for CDM \citep{Veltmaat2016, Schive2015, Hopkins2018, Mocz2015, Nori2018}. However, the fluid approach has its own set of limitations, notably its inability to accurately resolve interference patterns in regions where the quantum pressure becomes ill-defined due to vanishing densities \citep{Li2018, Zhang2019}.

In this paper, we develop a hybrid AMR scheme that integrates the fluid formulation of the Schrödinger-Poisson equations on coarser grids for large scales with an accurate algorithm to solve the wave formulation of the Schrödinger-Poisson equations on refined grids for small scales. This approach combines the advantages of both methods: The larger-scale grids do not require resolving the de Broglie wavelength, while the wave formulation on smaller scales accurately depicts interference phenomena. Such a hybrid code holds the promise of facilitating zoom-in simulations \revtext{capturing the dynamics of FDM across all relevant scales \citep{Liao2024, Chan2025, Chiu2025}. Here, zoom-in simulations refer to simulations that concentrate most computational resources on a specific region of interest (e.g. the Lagrangian volume identified from an N-body simulation with the same initial conditions). This is achieved by increasing the resolution within a sub-volume of the entire simulation box while maintaining a coarser resolution elsewhere.}

Pioneering efforts in this direction have been made by \textcite{Veltmaat2018, Schwabe2021}, who developed hybrid codes integrating a Lagrangian N-body solver for large scales with a finite difference wave solver equipped with AMR for smaller scales. In these works, the wave function is reconstructed by evolving the phase information of individual N-body particles and summing their information with appropriate interpolation kernels. Both methods can arguably reproduce the generic and statistical properties of FDM halos, but may fail to describe fine-grained halo properties such as granule, soliton, and subhalo dynamics because the reconstruction of the wave function and the boundary conditions are only approximate. In the work by \textcite{Veltmaat2018}, this is because the initial and boundary conditions of the Schrödinger domain are provided by a classical wave function that does not account for interference effects. In subsequent work by \textcite{Schwabe2021}, the reconstruction of the wave function is improved via a WKB-like Gaussian beam method with fixed amplitudes. It uses the particle information in the center of a pre-selected halo to construct a wave function with the statistically correct interference pattern. Yet, the method still ignores the quantum pressure term. In addition, the authors note that assuming fixed amplitudes for Gaussian beams overlooks the second-order spatial derivatives of the potential, necessitating a numerical verification to confirm that the respective contributions of the potential to the velocity dispersion in halos and filaments cancel out statistically.

In contrast, the method presented in this work does not require approximations to convert between the fluid and wave representations of the wave function, yielding results that are in theory point-wise comparable to those obtained from wave-only simulations of the Schrödinger equation. Moreover, the fluid scheme presented here incorporates the quantum pressure term and can therefore be applied to the quasi-nonlinear regime as long as vortices have not formed. Therefore, this work offers a more universal approach for simulating the Schrödinger equation, making it potentially suitable for studying generic and statistical as well as fine-grained properties of FDM halos.

The plan of this paper is as follows: In Section \ref{sec:methods}, we briefly review the Schrödinger-Poisson equation and the Madelung transform. In Section \ref{sec:algorithm}, we describe the proposed algorithms, including phase matching at fluid-wave boundaries, a fluid solver for the Hamilton-Jacobi-Madelung equations, a local pseudospectral wave solver for the Schrödinger equation, grid refinement strategies for both the fluid and wave solvers, a new interpolation algorithm, as well as the integration of these building blocks into the \texttt{GAMER} code. In Section \ref{sec:results}, we demonstrate the advantages and disadvantages of the proposed algorithms through a series of tests without gravity, tests of the linear power spectrum evolution, and a hybrid cosmological simulation.

\section{Governing Equations} \label{sec:methods}

\begin{figure*}[ht]
\centering
\includegraphics[width=1.0\textwidth]{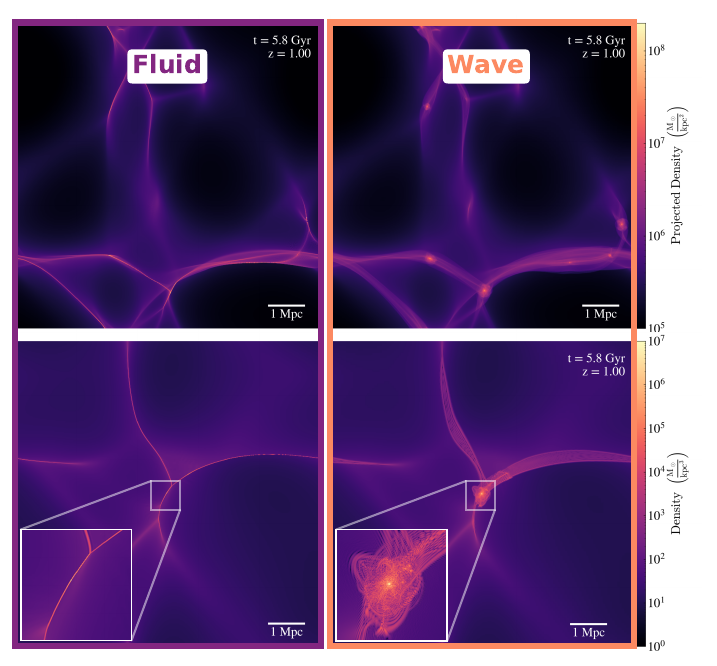}
\caption{Mismatch between the Hamilton-Jacobi-Madelung and Schrödinger equations in a cosmological simulation of the FDM model. Left: AMR simulation using the Hamilton-Jacobi-Madelung equations \eqref{eq:HamiltonJacobiMadelungContinuity} and \eqref{eq:HamiltonJacobiMadelungPhase}. Right: AMR simulation solving Eq. \eqref{eq:Schroedinger} in refined grid regions and Eqs. \eqref{eq:HamiltonJacobiMadelungContinuity} and \eqref{eq:HamiltonJacobiMadelungPhase} in coarser grid regions. Top: Density projection. Bottom: Density slice through a halo. The Hamilton-Jacobi-Madelung equations fail to describe interference patterns correctly. Instead, structures collapse into points and are not stabilized against collapse by the quantum pressure term. In contrast, the Schrödinger equation produces intricate interference patterns and an extended halo. Note that the total mass depicted in the density slices using the fluid and wave approaches disagrees because the fluid-only simulation generates high-density spots that are misaligned with the slices shown here.}
\label{fig:madelung_schroedinger_mismatch}
\end{figure*}

In this section, we introduce the fundamental equations relevant to the hybrid scheme. The Schrödinger-Poisson equations in comoving coordinates describe the non-relativistic approximation of the evolution of the wave function in the FDM model \citep{Marsh2016}:
\begin{align}
\label{eq:Schroedinger}
    i\hbar\left(\partial_t \psi + \frac{3}{2} H \psi\right) &= \left(-\frac{\hbar^2}{2m a^2}{\nabla}^2 + m \Phi\right) \psi, \\
    {\nabla}^2 \Phi &= 4\pi G a^2 (m|\psi|^2 - \rho_0(t)),
\end{align}
where $m$ is the FDM particle mass, $\hbar$ is the reduced Planck constant, $H$ is the Hubble parameter, $a$ is the scale factor, $G$ is the gravitational constant, $\rho_0(t)$ represents the background density, and $\Phi$ is the gravitational potential. They can be recast into a hydrodynamical form via the substitution 
\begin{equation}
    \label{eq:MadelungTransform}
    \reveq{\psi = \sqrt{\frac{\rho}{m}} e^{i S/\hbar}}
\end{equation}
for the real density field $\rho(\bm{x}, t)$ and the real phase field $S(\bm{x}, t)$. Furthermore, we simplify Eq. \eqref{eq:Schroedinger} via the transformations $\mathrm{d}t' = a^{-2} \mathrm{d}t$, $\rho' = a^{3} \rho$, and $\Phi' = a^2 \Phi$. Hereafter, ($\mathrm{d}t$, $\rho$, $\Phi$) refer to the variables ($\mathrm{d}t'$, $\rho'$, $\Phi'$) whenever adopting comoving coordinates. In addition, we normalize the comoving mass density $\rho$ to the comoving background density $\rho_{0}$. This leads to the following form of the Hamilton-Jacobi-Madelung equations:
\begin{align}
    \reveq{m \partial_t   \rho +  \bm{\nabla} \cdot \left(\rho \bm{\nabla} S\right)  = 0,}\label{eq:HamiltonJacobiMadelungContinuity}\\
    \reveq{m \partial_t S +  \frac{1}{2}\left(\bm{\nabla} S\right)^2 + m^2 \Phi - \frac{1}{2} \hbar^2 \frac{{\nabla}^2 \sqrt{\rho}}{\sqrt{\rho}} = 0,} \label{eq:HamiltonJacobiMadelungPhase}\\
    {\nabla}^2 \Phi - 4\pi G a \left(\rho - 1\right)  = 0.\label{eq:PoissonEquation}
\end{align}

By taking another gradient of Eq. \eqref{eq:HamiltonJacobiMadelungPhase} and defining $\reveq{\bm{v} = \bm{\nabla} S/m}$, one obtains the Madelung equations for frictionless, compressible flow in a gravitational potential modified by the quantum pressure term ${\nabla}^2\sqrt{\rho}/\sqrt{\rho}$:
\begin{align}
    \partial_t   \rho +  \bm{\nabla} \cdot \left(\rho \bm{v}\right)  = 0,\label{eq:MadelungContinuity}\\
    \partial_t \bm{v} +  \bm{v} \cdot \bm{\nabla} \bm{v} +  \bm{\nabla} \Phi - \frac{1}{2} \frac{\hbar^2}{m^2} \bm{\nabla} \frac{{\nabla}^2 \sqrt{\rho}}{\sqrt{\rho}} = 0.\label{eq:MadelungPhase}
\end{align}
On large scales, the quantum pressure term vanishes, and one recovers the pressureless, inviscid ideal fluid equations approximately applicable to CDM before shell crossing \citep{Bernardeau2001}. 

\revtext{One might wonder why one would replace a system of individually linear equations with a system that includes nonlinear terms.} One advantage lies in the applicability of hydrodynamics schemes that have less stringent spatial resolution requirements than Schrödinger-Poisson solvers. Despite its usefulness, the Madelung formulation assumes non-vanishing density $\rho$, a condition not always met in wave dynamics due to phenomena like destructive interference. In particular, the Madelung formulation inevitably fails in dark matter halos and filaments where vortices are ubiquitous. This limitation necessitates careful consideration of the applicability of the fluid formulation, particularly when interfacing with standard hydrodynamics codes, as discussed by \citet{Li2018, Zhang2019}.

Fig. \ref{fig:madelung_schroedinger_mismatch} highlights the mismatch between cosmological simulations relying on the Hamilton-Jacobi-Madelung equations and the Schrödinger equations.  It shows a cosmological simulation of the FDM model with $m = 2\times 10^{-23}$ eV in a simulation cube with a comoving side length $L=5.6$ Mpc/h at $z = 1$. We use comoving distances for cosmological simulations throughout the text. Both simulations use the \texttt{GAMER} code, with initial conditions generated using \texttt{axionCAMB} \citep{axionCAMB} and \texttt{MUSIC} \citep{Hahn2011} at $z=100$. The figure shows that the Hamilton-Jacobi-Madelung equations accurately capture the density distribution on large scales but produce incorrect density fields on small scales in regions of destructive interference, highlighting the need for a hybrid approach. 

\section{Algorithm} \label{sec:algorithm}
In this section, we present the components of a hybrid algorithm that seamlessly integrates a fluid solver on large scales with a wave solver in regions of strong interference. Firstly, the integration of the Madelung and Schrödinger equations requires precise matching of the wave function at the fluid-wave boundary. This is why we choose to evolve the Hamilton-Jacobi-Madelung equations \eqref{eq:HamiltonJacobiMadelungContinuity} and \eqref{eq:HamiltonJacobiMadelungPhase} instead of the Madelung equations \eqref{eq:MadelungContinuity} and \eqref{eq:MadelungPhase}, as explained in Section \ref{subsec:boundary_matching}.

In Section \ref{subsec:hamilton_jacobi}, we introduce a second-order spatially accurate algorithm for the Hamilton-Jacobi-Madelung equations. It combines a Monotonic Upstream-centered Scheme for Conservation Laws (MUSCL) for evolving the continuity equation and an upwind algorithm for the Hamilton-Jacobi equation with a third-order Runge-Kutta method. 

Next, we delve deeper into wave solvers in Section \ref{subsec:local_spectral_solver}. When integrating a wave solver with a fluid solver, one significant challenge is that the wave solver requires substantially higher resolution than the fluid solver at the wave-fluid interface. This is because finite difference schemes for wave equations typically require about $10-20$ points per wavelength to achieve sufficient accuracy (also see Rule-of-Thumb 4 in \citet{Boyd_2001}). We partially mitigate this requirement by introducing a $13$th-order spatially accurate local pseudospectral algorithm based on Fourier continuations with Gram polynomials for the linear Schrödinger equation. This algorithm has lower resolution requirements compared to conventional finite difference algorithms.

Building on this, we present an accurate interpolation algorithm using Fourier continuations with Gram polynomials in Section \ref{subsec:local_spectral_interpolation}. It interpolates the density and phase fields in smooth regions of the solution and switches to the real and imaginary parts of the wave function near vortices.

With a view to their implementation in AMR codes, the fluid and wave solvers are enhanced by two mesh refinement strategies: the first refinement strategy, presented in Section \ref{subsec:madelung_refinement}, applies to the fluid solver. It is based on the quantum pressure term and the second derivative of the phase to ensure accurate switching between the fluid and wave algorithms. The second refinement strategy, presented in Section \ref{subsec:spectral_refinement}, applies to the wave solver. It is based on the magnitude of the coefficients of the polynomial expansion of the wave function to ensure that the wave function is well-resolved.

Finally, we explain the integration of the hybrid scheme into the AMR code \texttt{GAMER} in Section \ref{subsec:integration_into_gamer}. 

\subsection{Boundary Matching Problem}
\label{subsec:boundary_matching}
In this section, we address the integration of fluid and wave algorithms, which necessitates the precise reconstruction of the wave function from the Hamilton-Jacobi-Madelung formulation as well as the density and phase fields from the Schrödinger-Poisson equations, as outlined in Eq. \eqref{eq:MadelungTransform}. This task is referred to as the boundary matching problem in the following.

\begin{figure*}[ht!]
\centering
\includegraphics[width=1.0\textwidth]{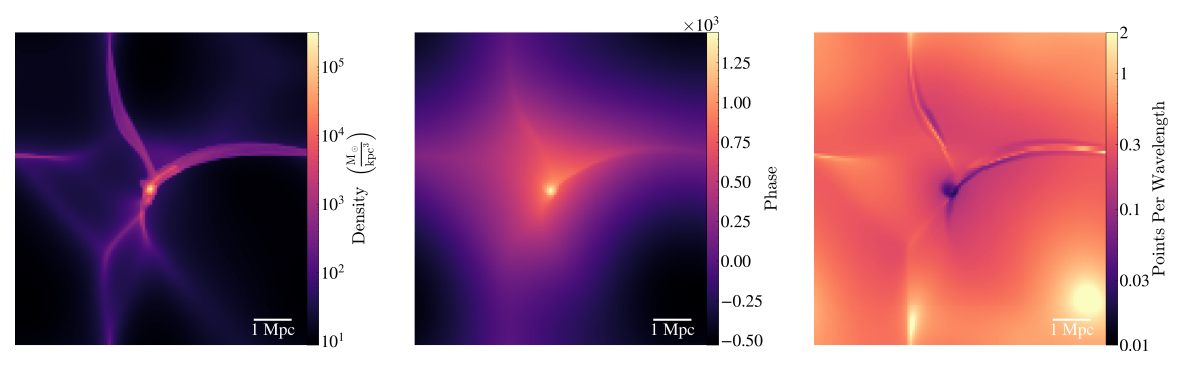}
\caption{Slices of the density field (left), the phase field (center), and the number of grid points per de Broglie wavelength (right) on the root level of the hybrid cosmological simulation shown in Fig. \ref{fig:madelung_schroedinger_mismatch}. The root level has $N=128^3$ grid points. The number of grid points per wavelength is estimated as $\reveq{2\pi\hbar/(\Delta x |\bm{\nabla} S|)}$, where $\Delta x$ is the grid spacing. Crucially, the right figure demonstrates that the Hamilton-Jacobi-Madelung equations allow for phase differences much greater than $2\pi$ between neighboring points.}
\label{fig:points_per_wavelength}
\end{figure*}

When evolving the density and velocity fields using Eqs. \eqref{eq:MadelungContinuity} and \eqref{eq:MadelungPhase}, reconstructing the real and imaginary parts of the wave function presents a challenge: the solution is not unique. Whether one calculates a line integral of the velocity field or solves a Poisson equation for the divergence of the velocity field, the resulting phase field $S$ is determined only up to an integration constant. This constant is time-dependent and influences the evolution of the wave function. To resolve this, one can track the temporal evolution of the phase field by solving the Hamilton-Jacobi 
equation in addition to evolving the Madelung equations. The phase field is then updated using the given density and velocity fields. This is similar to the approach taken by the N-body wave solver hybrid schemes presented in \textcite{Veltmaat2018, Schwabe2021}. They initially assign phases to individual N-body particles by solving the Poisson equation for the divergence of the velocity field. They then use the Hamilton-Jacobi equation without the quantum pressure term to evolve these phases in time using the particle velocities and densities given by the N-body solver. 

In this work, we opt against evolving the density and velocity fields and instead evolve the density and phase fields using the Hamilton-Jacobi-Madelung equations \eqref{eq:HamiltonJacobiMadelungContinuity} and \eqref{eq:HamiltonJacobiMadelungPhase}. This method facilitates the direct reconstruction of the wave function $\psi$ without incurring additional computational costs. Yet, for the hybrid scheme presented here, evolving the phase introduces the \emph{reverse} boundary matching problem: uniquely reconstructing the phase field from the wave function. This is necessary because the wave and the fluid subdomains need to mutually exchange information in a consistent simulation, e.g. by providing boundary conditions for one another. While deriving the velocity $\bm{v}$ from the wave function is straightforward, the phase is only defined up to a multiple of $2\pi$:
\begin{equation}
\label{eq:PhaseFromWave}
    \reveq{S/\hbar = \atantwo\left(\Im(\psi),\Re(\psi)\right) + 2 n \pi,}
\end{equation}
where $\Re(\psi)$ and $\Im(\psi)$ are the real and imaginary parts of the wave function $\psi$.  
The constant $n \in \mathbb{Z}$ can be uniquely determined by ensuring phase continuity at the grid interface as long as the phase difference between adjacent grid points does not exceed $2\pi$.  This requirement effectively means that the de Broglie wavelength of the wave function must be resolved at the interfaces between the fluid and wave formulations.

Fig. \ref{fig:points_per_wavelength} highlights that the Hamilton-Jacobi-Madelung equations support much lower resolutions in a typical hybrid cosmological simulation. It shows the root AMR level of the simulation highlighted in Fig. \ref{fig:madelung_schroedinger_mismatch}, corresponding to a resolution of $128^3$ points. The phase difference between adjacent grid points exceeds $2\pi$ in most of the simulation domain. This necessitates sufficiently increasing the grid resolution before switching to the wave formulation. Section \ref{subsec:madelung_refinement} introduces the refinement criterion for this purpose. Section \ref{subsec:integration_into_gamer} explains why and how we determine the minimum grid resolution before switching to the wave formulation.  

\begin{figure*}[ht!]
\centering
\includegraphics[width=\textwidth]{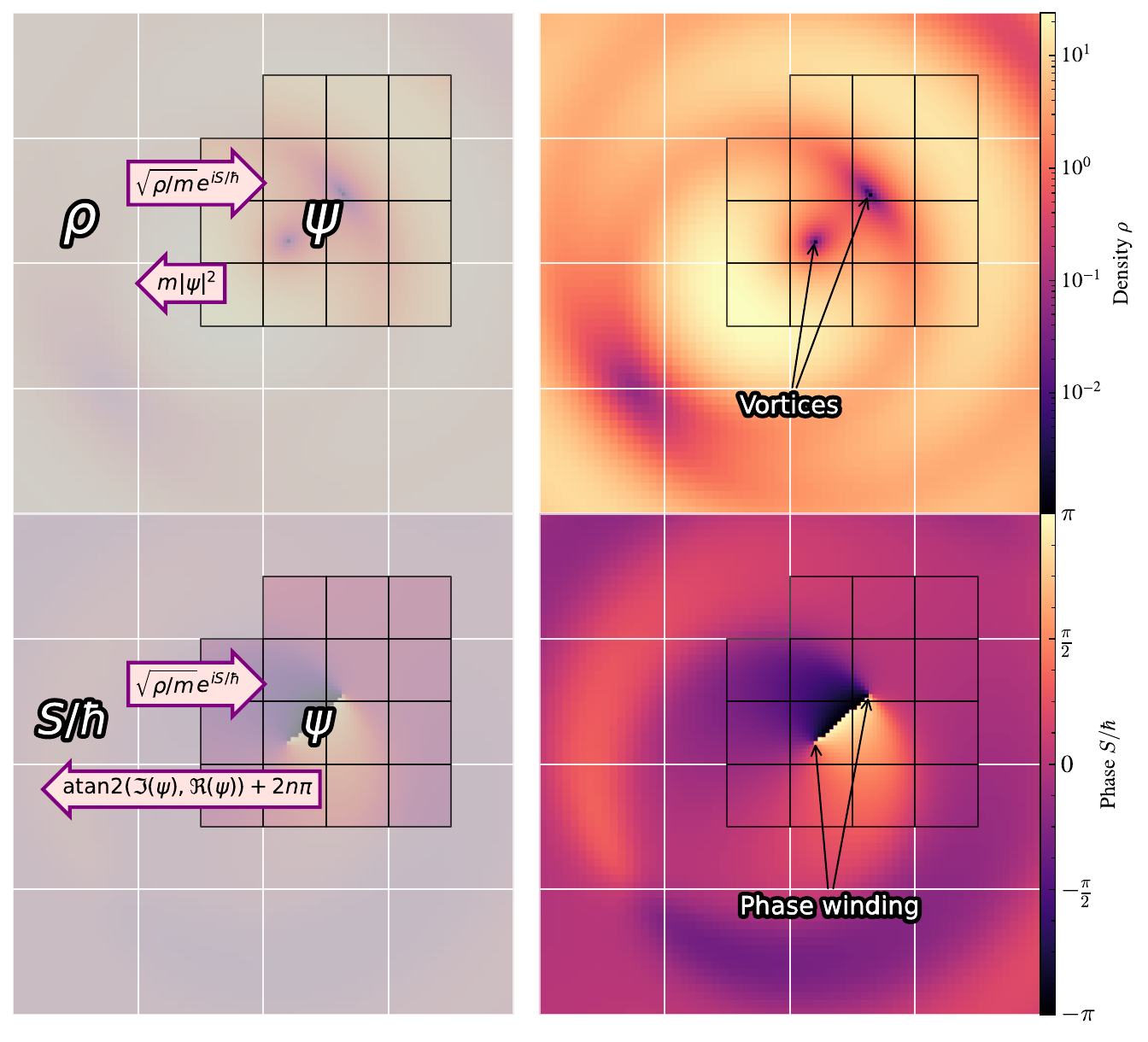}
\caption{Illustration of boundary matching for the co-rotating vortex pair described by Eq. \eqref{eq:VortexPairRotating}. The Hamilton-Jacobi-Madelung equations are applied in regions represented by the density $\rho$ and the phase $\reveq{S/\hbar}$, while the wave equation is solved in the subdomain denoted by the wave function $\psi$. The two sets of variables $(\rho, S)$ and $(\Re(\psi), \Im{(\psi)})$ can be uniquely converted into one another for the fluid-wave boundary shown here. \revtext{However, this is not the case if the fluid-wave boundary crosses regions of vanishing density. If such regions are isolated, they correspond to vortices, which are characterized by a nonzero circulation of the velocity field.} This non-zero circulation translates into phase windings where the phase acquired by $\psi$ along a closed cycle around the vortex is $\reveq{2 \pi}$ times a non-zero integer. The two vortices are connected by a phase discontinuity that can be uniquely unwrapped as long as the de Broglie wavelength is well resolved \revtexttwo{(manifesting as a phase jump of $2 \pi$ perpendicular to the line connecting the vortices)}. However, the phase field exhibits a distinct $\pi$ discontinuity directly at the vortices that is challenging to handle \revtexttwo{(occurring along the line connecting the vortices; see also Fig. \ref{fig:singular_gauge})}.}
\label{fig:boundary_matching}
\end{figure*}

Fig. \ref{fig:boundary_matching} demonstrates the reverse boundary matching process adopted in the hybrid scheme using the example of a co-rotating vortex pair that solves the linear Schrödinger equation \revtext{without gravity}:
\begin{equation}
\label{eq:VortexPairRotating}
\reveq{\psi(R, \varphi)} = \rho_{bg} - A J_1\left(R\sqrt{2 \omega m/\hbar}\right) \exp\left(i (\varphi - \omega t)\right),
\end{equation}
where $J_1$ is the Bessel function of the first kind, $R$ is the cylindrical radius, $\varphi$ is the polar angle, and we choose the parameters $\rho_{bg} = 2$, $A = 5$, $\omega = 90$, $m/\hbar=1$,  and $t = 8\times10^{-3}$ \revtext{assuming arbitrary code units}. The simulation employs a quadratic grid with a root-level resolution of $N=64^2$ points, boundary conditions provided by Eq. \eqref{eq:VortexPairRotating}, and one refinement level using a wave solver with the refinement controlled by the Madelung refinement criterion introduced in Section \ref{subsec:madelung_refinement}. Note that the $2 \pi$-discontinuities of the phase field connecting the vortex pair do not represent a problem for the Hamilton-Jacobi-Madelung equations as the phase can be uniquely unwrapped as long as the de Broglie wavelength is well resolved. However, the phase field also exhibits a $\pi$-discontinuity at the vortices that causes the Hamilton-Jacobi-Madelung equations to fail. As a result, the fluid-wave boundary must be positioned away from the vortex pair to avoid this issue. Section \ref{subsec:local_spectral_interpolation} and Fig. \ref{fig:singular_gauge} analyze this problem in more detail.

\subsection{Hamilton-Jacobi-Madelung Equations}
\label{subsec:hamilton_jacobi}
The Hamilton-Jacobi-Madelung equations consist of the continuity equation \eqref{eq:HamiltonJacobiMadelungContinuity} coupled with the Hamilton-Jacobi equation \eqref{eq:HamiltonJacobiMadelungPhase}, including additional quantum pressure and potential terms. 

\subsubsection{Operator Splitting}
Before discretizing the Hamilton-Jacobi-Madelung equations, we first address how to handle their coupling with the Poisson equation \eqref{eq:PoissonEquation} through the gravitational potential $\Phi$. The Schrödinger and Poisson equations can be separated using operator splitting. The time evolution operator $\hat{U}_{K+V}(\Delta t)$ of the Schrödinger-Poisson equations can be expressed as a series of applications of the so-called kick operator $\hat{U}_V(\Delta t) = \exp\left(-i m \Phi \Delta t / \hbar \right)$, corresponding to the potential term of the Schrödinger equation, and the so-called drift operator $\hat{U}_K(\Delta t) = \exp\left( -i\hbar/(2m) k^2 \Delta t\right)$, where $k$ is the momentum in Fourier space and the expression corresponds to the kinetic term of the Schrödinger equation. A first-order approximation of the full time evolution operator $\hat{U}_{K+V}(\Delta t)$ is given by the kick-drift scheme: 
\begin{equation}
\label{eq:kick_drift}
    \hat{U}_{K+V}(\Delta t) = \hat{U}_V\left(\Delta t\right) \circ \hat{U}_K\left(\Delta t\right) + \mathcal{O}(\Delta t^2).
\end{equation}

In the context of the Hamilton-Jacobi-Madelung equations, the kinetic term in the Schrödinger equation reflects the combined evolution of the continuity and Hamilton-Jacobi equations, including the quantum pressure but excluding the potential term. In theory, an initialization of the algorithm with an initial kick by half a time step would make the time evolution operator Eq. \eqref{eq:kick_drift} equivalent to a kick-drift-kick scheme and therefore second-order accurate in time. In practice, including the initial kick can be understood as modifying the initial condition by a perturbation of order $\Delta t$. Hence, both the kick-drift and the kick-drift-kick schemes will behave similarly as long as a simulation does not sensitively depend on the initial conditions.

Having explained how to account for the potential operator, we focus on the discretization of the drift operator $\hat{U}_K(\Delta t)$  for the Hamilton-Jacobi-Madelung equations in the following sections. The discussion will be limited to the one-dimensional case as the generalization to more dimensions using a directional splitting method is straightforward: One applies the drift operator to one dimension at a time.

\subsubsection{Continuity Equation}
Due to the dispersive nature of the quantum pressure term, the continuity equation does not develop shocks and the Hamilton-Jacobi equation does not develop kinks. However, the phase field is discontinuous at vortices. If the Hamilton-Jacobi-Madelung equations are used to evolve a solution containing vortices, it is important to ensure that the scheme remains well-behaved. Therefore, we discretize the continuity equation in space using a MUSCL-type scheme as a conservative finite difference method in conjunction with the van Albada limiter \citep{Leer1979, Albada1997}. 
We use a second-order linear subgrid model, meaning that the density within each cell is approximated as a linear function for the density update. In contrast, a scheme assuming constant densities within each cell would be first-order in space and a parabolic subgrid model would result in a third-order scheme. The resulting spatial discretization is expressed in flux form as 
\begin{equation} 
\partial_t \rho_i(t) = (F_{i-\frac{1}{2}} - F_{i+\frac{1}{2}})/\Delta x,
\end{equation} with the mass fluxes
\begin{equation}
\begin{split}
\label{eq:conservationlawfluxform}
     F_{i+\frac{1}{2}} &= \max(0, v_{i+\frac{1}{2}}) \rho_{i} + \min(v_{i+\frac{1}{2}}, 0) \rho_{i+1} \\&+ \frac{1}{2} |v_{i+\frac{1}{2}}| \left(1 - |v_{i+\frac{1}{2}}|\Delta t/\Delta x\right) \phi(r_{i+\frac{1}{2}}) (\rho_{i+1} - \rho_{i}),
\end{split}
\end{equation}
where $v_{i\pm\frac{1}{2}}$ are the velocities at the boundaries of cell $i$
\begin{equation}
\label{eq:cell_boundary_velocity}
v_{i\pm\frac{1}{2}} =(\pm S_{i\pm1} \mp S_i)/\Delta x,     
\end{equation}
and $\phi(r)$ is the van Albada limiter
\begin{equation}
    \phi(r) = (r^2 + r)/(1 + r^2)
\end{equation}
with the arguments
\begin{equation}
r_{i+\frac{1}{2}} \equiv
\begin{cases}
\begin{alignedat}{2}
&(\rho_i - \rho_{i-1})/(\rho_{i+1} - \rho_i),\quad &\mathrm{for}&\quad v_{i+\frac{1}{2}} \geq 0,\\
&(\rho_{i+2} - \rho_{i+1})/(\rho_{i+1} - \rho_i),\quad &\mathrm{for}&\quad v_{i+\frac{1}{2}} \leq 0.
\end{alignedat}
\end{cases}
\end{equation}
Here $\Delta x$ is the cell size, $\Delta t$ is the size of the time step, and we take $\hbar = m = 1$ for simplicity. 
The left and right fluxes agree between neighboring cells, i.e. the left flux of the $i$-th cell $F_{i-\frac{1}{2}}$ agrees with the right flux of the $(i-1)$-th cell $ F_{(i-1)+\frac{1}{2}}$. The resulting scheme is second-order accurate in space \citep{Nishikawa2020} and conserves mass.

\subsubsection{Hamilton-Jacobi Equation}
The one-dimensional Hamilton-Jacobi equation is discretized using a different approach. The convection term is discretized in space using the Sethian-Osher flux \citep{Shu2007}:
\begin{equation}
    \left(\nabla S\right)^2_i = \min\left(v_{i+\frac{1}{2}}, 0\right)^2 + \max\left(v_{i-\frac{1}{2}}, 0\right)^2,
\end{equation}
with the third-order cell boundary velocities 
\begin{equation}
    v_{i\pm\frac{1}{2}} = \left(\mp 2 S_{i\mp1} \mp 3 S_i \pm 6 S_{i\pm1} \mp S_{i\pm2}\right)/(6 \Delta x).     
\end{equation}
The quantum pressure term is discretized in the form
\begin{equation}
\label{eq:QPDiscretization}
    \frac{\nabla^2 \sqrt{\rho}}{\sqrt{\rho}} = \frac{1}{2} \nabla^2 \log(\rho) + \frac{1}{4} \left(\nabla \log(\rho)\right)^2,
\end{equation}
using second-order central finite differences in space for both the gradient and Laplacian operators. This leads to the expression
\begin{equation}
\begin{split}
\left(\frac{\nabla^2 \sqrt{\rho}}{\sqrt{\rho}}\right)_i &= \frac{1}{2}\frac{\log(\rho_{i+1}) - 2 \log(\rho_i) + \log(\rho_{i-1})}{\Delta x^2}\\
&+ \frac{1}{4} \left(\frac{\log(\rho_{i+1}) - \log(\rho_{i-1})}{2\Delta x}\right)^2.
\end{split}
\end{equation}
The semi-discrete form of the Hamilton-Jacobi equation reads
\begin{equation}
\label{eq:HJScheme}
\partial_t S_i(t) =  - \frac{1}{2} (\nabla S)^2_i  + \frac{1}{2}\left(\frac{\nabla^2 \sqrt{\rho}}{\sqrt{\rho}}\right)_i.
\end{equation}
This scheme does not conserve momentum and we are unaware of any momentum-conserving scheme to evolve the phase field directly.

\subsubsection{Time discretization}
Time discretization is performed using a strong stability-preserving third-order Runge-Kutta method suggested by \citet{Shu2007}. Defining $u^n = \begin{pmatrix} \rho^n & S^n \end{pmatrix}^\intercal$ and $L(u^n, t^n) = \begin{pmatrix} \partial_t \rho^n & \partial_t S^n \end{pmatrix}^\intercal$, where the \revtext{superscript} $n$ denotes the time step, it reads
\begin{equation}
\label{eq:RK3}
\begin{split}
    u^{(1)} &= u^n + \Delta t L(u^n, t^n),\\
    u^{(2)} &= \frac{3}{4} u^n + \frac{1}{4} u^{(1)} + \frac{1}{4} \Delta t L(u^{(1)}, t^n + \Delta t),\\
    u^{n+1} &= \frac{1}{3} u^n + \frac{2}{3} u^{(2)} + \frac{2}{3} \Delta t L(u^{(2)}, t^n + \frac{1}{2} \Delta t).
\end{split}
\end{equation}
Although using a second-order Runge-Kutta method allows for a smaller ghost region of $4$ instead of $6$ points and remains stable, it requires a more stringent Courant–Friedrichs–Lewy (CFL) stability condition. We are unaware of any convergence guarantees for the resulting scheme and must rely on numerical experimentation to assess its properties. Consequently, the time steps are chosen to satisfy the CFL condition:
\begin{equation}
\label{eq:fluid_cfl}
    \Delta t \leq  \min\left[C_D \frac{m}{\hbar} \Delta x^2, C_K \Delta x \frac{m}{\reveq{2 \sum_{j=1}^3 |\partial_j S|}}\right],
\end{equation}
where $C_D = 0.2$ and $C_K = 1.0$, with the summation of the partial derivatives $\partial_j S$ taken over three spatial dimensions.

\subsection{Local Pseudospectral Solver}
\label{subsec:local_spectral_solver}
This section presents an accurate method for solving the Schrödinger equation: a local pseudospectral method using Fourier continuations with discrete Gram polynomials, the so-called FC-Gram algorithm. \revtexttwo{The discrete Gram polynomials $t_n(x)$ of degree $n$ used in the following are a set of orthogonal polynomials defined with respect to a discrete scalar product $(t_m, t_n) \equiv \sum_{i=0}^{N-1} t_m(x_i)t_n(x_i)$ over $N$ equispaced points $x_i$. They satisfy the orthogonality condition $(t_m, t_n) = 0$ for $m \neq n$ with $m, n = 0, \hdots, N-1$, which makes them particularly useful for approximation problems.  Like all classical orthogonal polynomials, the discrete Gram polynomials satisfy a three-term recurrence relation:
$(n+1)t_{n+1}(x) = (2n+1)(2x-N+1)t_n(x) - n(N^2-n^2)t_{n-1}(x)$
with initial conditions $t_0(x) = 1$ and $t_1(x) = 2x - N + 1$. The squared norm can be computed recursively as
$H_n^2 = (2n-1)/(2n+1)(N^2-n^2)H_{n-1}^2$ with the initial condition $H_0^2 = N$. To obtain orthonormal polynomials $T_n(x)$ that satisfy the normalization condition $(T_m, T_m) = 1$ for $m = 0, \dots, N-1$, one can simply normalize using $T_n(x) = t_n(x)/H_n$.}

\revtext{The FC-Gram algorithm marks an improvement over conventional finite difference methods for solving the Schrödinger equation at a comparable computational cost.}
For periodic boundary conditions, a pseudospectral discretization using discrete Fourier transforms is the algorithm of choice for the solution of the Schrödinger equation \citep{Woo2009, Mocz2017, May2021}. It discretizes the kinetic operator of the Schrödinger equation in Fourier space and achieves spectral convergence, that is, faster than any polynomial for smooth $\psi(x, t)$.
However, FDM AMR codes such as \texttt{AxioNyx} \citep{Schwabe2020}, \texttt{SCALAR} \citep{Mina2020}, and \texttt{GAMER} divide the simulation domain into subdomains with different mesh resolutions. In these subdomains, the wave function is not necessarily periodic, motivating the aforementioned codes to use finite difference methods for the linear Schrödinger equation in refined regions. Although \texttt{SCALAR} and \texttt{GAMER} also implement an accurate pseudospectral method on the root level of the simulation domain, we find that the accuracy of AMR simulations is ultimately limited by the algorithm employed to evolve the wave function on refined levels.

\subsubsection{Overcoming the Gibbs Phenomenon}
\begin{figure*}[ht!]
\centering
\includegraphics[width=\textwidth]{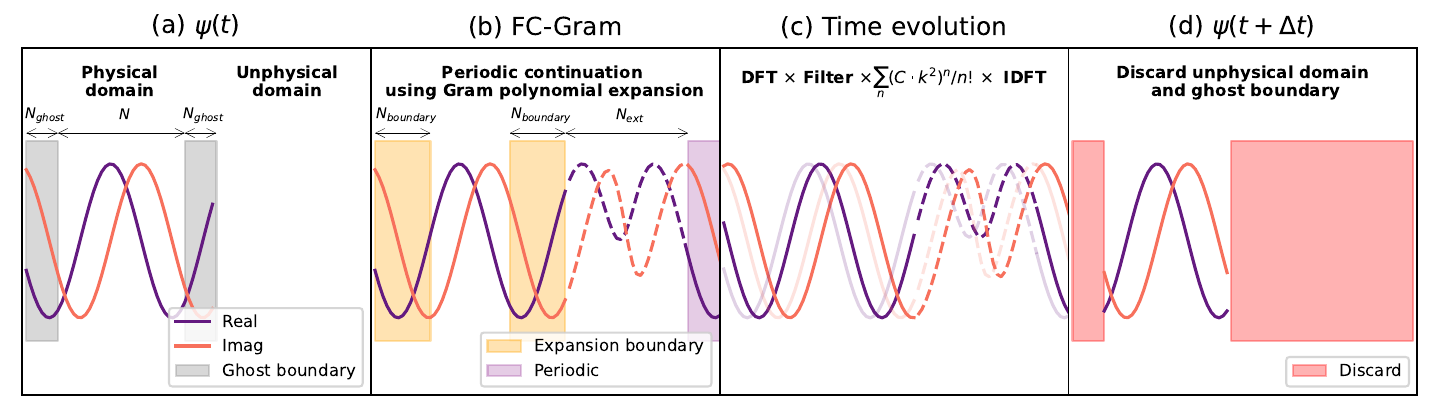}
\caption{Overview of the pseudospectral FC-Gram solver for the free Schrödinger equation. The physical part and the periodic extension of the real and imaginary parts of the wave function are shown through solid and dashed lines, respectively. The algorithm consists of three steps: It computes an extension of the physical data that is periodic in the extended domain (b), evolves the wave function in Fourier space (c), and finally discards both the unphysical domain and the ghost regions (d). See Section \ref{sec:GFextension} for more details.}

\label{fig:gramfe_fft}
\end{figure*}
Evolving the non-periodic data in the AMR subdomains using a Fourier transform leads to the Gibbs phenomenon\revtext{, which arises due to the inherent assumption of periodicity in Fourier-based methods. This results in spurious oscillations near discontinuities or sharp gradients, preventing uniform convergence and leading to slow numerical convergence.} Fortunately, this slow convergence can be overcome by a wide range of methods that attempt to recover fast convergence when reconstructing non-periodic data on a uniform grid with $N$ points \citep{Boyd_2009, Boyd_2011}.
These methods include, among others, subtraction methods \citep{Skoellermo1975}, filters and mollifiers \citep{Tadmor2007}, Local Fourier Basis method \citep{Israeli1993}, Gegenbauer methods \citep{Gottlieb1992}, Inverse Polynomial Reconstructions \citep{Jung2004}, Polynomial Frame Approximations \citep{Adcock2023} and Singular Value Decomposition (SVD) extensions \citep{Boyd2002}. Some of these methods have been successfully used to construct stable, highly accurate partial differential equation (PDE) solvers on uniform grids \citep{Israeli1993, Averbuch1998, Braverman2004, Israeli2007, Albin2011}.

The above-listed methods exhibit algebraic convergence (e.g. subtraction methods, Local Fourier Basis method) to exponential convergence (e.g., Polynomial Frame Approximation with regularization). But numerical ill-conditioning is a serious problem for all of the above methods. \revtext{
Partly, this is because polynomial-based methods must also contend with the Runge phenomenon, which manifests as large oscillations near boundaries when using high-degree polynomials for interpolation on equispaced points. Relatedly, there is strong numerical evidence that the condition number of interpolation matrices in many of the above-listed methods grows rapidly with the number of grid points $N$ \citep{Boyd_2009}.} For instance, Gegenbauer methods and Inverse Polynomial Reconstructions exhibit large errors close to the domain boundaries if the regularization is not suitably chosen. This is of course undesirable for a stable, versatile PDE solver. Moreover, \cite{Platte2011} showed that equally spaced samples can achieve root-exponential convergence but not exponential convergence, whereas \cite{Adcock2023} found that a well-conditioned method, like Polynomial Frame Approximation, can achieve near-exponential error reduction to a user-defined tolerance. These numerical studies highlight the importance of striking a balance between convergence, stability, and numerical conditioning when designing a PDE solver.

It is important to note that in an AMR code, the convergence speed is limited by the size of the AMR subdomains unless there is global all-to-all communication between subdomains at every time step. Many AMR codes including \texttt{GAMER} avoid such global all-to-all communication to ensure high parallel efficiency. Without global all-to-all communication, the maximum spatial convergence that we can expect in an AMR code is algebraic and determined by the maximum polynomial order that an AMR subdomain of size $N$ supports. As a result, we compared the aforementioned approaches based on the number of points needed before the decay of the error sets in (i.e. resolution power) for a given small subdomain size, their numerical conditioning, and their convergence rates.

After extensive numerical experiments, we opted for the FC-Gram algorithm, first introduced by \cite{ Lyon_2010, Bruno_2010} with the improvements suggested by \cite{Albin2011}. The FC-Gram algorithm offers a computationally efficient method to compute an accurate periodic extension of non-periodic data, allowing for the application of accurate pseudospectral algorithms. It resolves the trade-off between convergence speed and numerical conditioning by abandoning geometric or sub-geometric convergence and settling for high-order algebraic convergence. Additionally, we found that the FC-Gram algorithm exhibited higher resolution power for the subdomain sizes relevant to \texttt{GAMER} compared to, for instance, the Inverse Polynomial Reconstruction method.  
 
\subsubsection{FC-Gram scheme for the Schrödinger Equation}
\label{sec:GFextension}
Fig. \ref{fig:gramfe_fft} visualizes how the FC-Gram algorithm can be used to build a solver for the linear Schrödinger equation. 
\begin{enumerate}[label=(\alph*)]
    \item \revtext{The algorithm begins with the physical wave function defined on the physical domain of size $N$. This domain is extended with left and right ghost regions, each of size $N_{ghost}$. These ghost regions serve as buffer zones to facilitate the exchange of physical wave functions between neighboring subdomains. In the context of the FC-Gram algorithm, the ghost regions ensure that artifacts in the unphysical domain introduced from the periodic extension process remain sufficiently distant from the physical data that is retained for each patch.}

    \item \revtext{At the boundaries of the physical domain, we expand the real and imaginary parts of the wave function into Gram polynomials. The expansion coefficients $a_n^{\text{left}}$, $a_n^{\text{right}}$ are computed as
    \begin{align}
        a_n^{\text{left}} &= \sum_{x \in \Omega_{\text{left}}} \psi(x)\, T_n(x),\\
        a_n^{\text{right}} &= \sum_{x \in \Omega_{\text{right}}} \psi(x)\, T_n(x),
    \end{align}
    where $T_n(x)$ are the orthonormal Gram polynomials of order $n=0,1,\dots,N_{\text{boundary}}-1$, and $\Omega_{\text{left/right}}$ denote boundary point sets of size $N_{\text{boundary}}$.} Using these coefficients, we then extend the wave function to an unphysical region of size $N_{ext}$, which, together with the physical domain, forms the total computational domain of size $N + 2 N_{ghost} + N_{ext}$. This extension is done by replacing the Gram polynomials with a set of functions that match the Gram polynomials within the physical domain and are periodic across the total computation domain.

    \item \revtext{The extended, periodic wave function is evolved in Fourier space by multiplication with a truncated Taylor expansion of the kinetic operator $\exp(-i \Delta t \hbar k^2/2m)$ with an additional exponential filter to ensure stability:
    \begin{equation}
         \exp\left(-\alpha (k/k_{max})^{2\beta}\right) \sum_{n=0}^{N_{taylor}} \frac{(C k^2)^n}{n!},
    \end{equation}
    with $C = -i \Delta t \hbar/2m$ and the filter parameters $\alpha$, $\beta$ and $k_{max} = \pi/\Delta x$ as suggested by \citet{Albin2011} (see Table \ref{tab:fcgramparams}).}

    \revtext{The Taylor truncation and exponential filtering are crucial in suppressing spurious high-frequency oscillations (known as spectral blocking), which could otherwise destabilize the numerical scheme \citep[p.~202]{Boyd_2001}. Specifically, higher-order terms in the Taylor expansion (i.e., larger values of $N_{taylor}$) represent the evolution of high-frequency (large-$k$) modes in the solution more accurately. Although enhanced accuracy may appear beneficial, these higher-order terms simultaneously amplify unwanted high-frequency oscillations introduced by the artificial periodic extensions. Hence, Taylor truncation effectively acts as a spectral filter that, in combination with the explicit exponential filtering, ensures numerical stability by limiting this amplification.}  

    \item After applying an inverse Fourier transform, the wave function within the unphysical domain and the ghost region is discarded. Note that, unlike a global Fourier method, this algorithm does not conserve mass. While the Fourier transform is unitary, mass is only conserved on the total computational domain. During the application of the kinetic operator, mass is exchanged between the physical and unphysical domains.
\end{enumerate}

For details on obtaining the periodic extensions of the Gram polynomials, please refer to \citet{Lyon2009}. In the following, we provide a brief overview of the algorithm. \revtext{The periodic functions are obtained by computing SVD Fourier continuations of the Gram polynomials for the FC-Gram algorithm. Let $\hat{f}(x)$, defined in $[0, \Theta]$, be the periodic extension of a Gram polynomial $f(x)$ defined in $[0, \chi]$, where $\Theta > \chi$.
$\hat{f}(x)$ allows for an accurate Fourier expansion as $\hat{f}(x) = \sum_l a_l e^{i 2 \pi  l x/\Theta}$. The crucial idea of SVD Fourier continuations is that the Fourier coefficients $a_l$ can be obtained by solving a linear optimization problem:
\begin{equation}
\begin{split}
\label{eq:SVD}
    \min_{a_l \forall l = 1, \hdots, g} \sum^{\Lambda}_{j=1} \reveq{\left|a_l M_{lj} - f_j\right|^2},
\end{split}
\end{equation}
where the points $x_j$ for $j=1,\hdots, \Lambda$ are collocation points in the physical domain $[0, \chi]$, $f_j \equiv f(x_j)$ is defined at the collocation points, and the matrix $M$ is defined as $M_{lj} = e^{i 2 \pi A_l x_j/\Theta}$ for $l=1, \hdots, g$ and $j=1, \hdots, \Lambda$ for a set of suitably chosen wave vectors $A_l$ (for instance, $A_l = l$, $A_l = 2l$ or $A_l = 2l-1$) corresponding to the Fourier coefficients $a_l$.}

By finding a minimum for the optimization problem, one ensures that the mismatch between $f(x)$ and its extension $\hat{f}(x)$ is small in the physical domain. The method derives its name from the fact that the optimization problem can be solved accurately using an SVD. \revtext{Importantly, the SVD is computed only once before runtime, ensuring that the expensive computation is not repeated. Once the SVD is precomputed, the \revtexttwo{FC-Gram} algorithm utilizes these precomputed extensions, which are stored either directly in the source code or in external files, making the runtime implementation highly efficient.}
In practice, we compute the SVD of the matrix $M$ with high floating-point precision with $256$ significant digits using Python's \texttt{mpmath} library \citep{mpmath}.

An additional complication arises from the fact that we compute two independent extensions for the left and right subdomain boundaries: How can they be put together? \cite{Lyon2009} proposes to compute even and odd extensions using only even and odd wave vectors, respectively. By taking linear combinations of the two, one obtains extensions that smoothly decay to zero. These extensions can then be combined to obtain a globally periodic extension. While the so-obtained extensions are unphysical, they reduce the Gibbs phenomenon by providing a smooth, periodic continuation of the non-periodic data that allows for a rapidly convergent Fourier expansion. 

From the above, it is clear that SVD continuations can also be directly applied to the input data. This raises the question: Why use Gram polynomials when a periodic extension can be computed directly? The main reason is that SVD continuations require over-collocation for high accuracy. This means that the number of collocation points $\Lambda$ should be greater than the number of wave vectors $g$. For input data with limited resolution, this fundamentally limits the accuracy of SVD continuations. The FC-Gram algorithm solves this problem by computing SVD continuations for analytically known basis functions. This is similar to first interpolating the input function and then computing the extension. The price to pay is that the smoothness of the Fourier continuation is limited by the quality of this interpolation: The fact that the extension is computed based on a polynomial expansion of limited order implies that the extension cannot be arbitrarily smooth at the subdomain boundary. An expansion boundary containing $N_{boundary}$ points supports polynomials of order $\leq N_{boundary} - 1$. This means that one can estimate at most $N_{boundary} - 1$ derivatives. Hence, the extension generally has up to $N_{boundary} - 1$ continuous derivatives. As a result, the Fourier coefficients of the FC-Gram continuation decay algebraically as $\mathcal{O}(\reveq{|k|}^{-N_{boundary}})$ at best \citep{Skoellermo1975}. In other words, an expansion boundary containing $N_{boundary}$ points leads to algebraic convergence of order $N_{boundary}$ (see \cite{Lyon_2010} and \cite{Anand2024} for a more detailed error analysis). Since we expand the input data in terms of polynomials, their maximum order is also limited by the Runge phenomenon. 

At first glance, this procedure may appear to offer little benefit. Why not opt for a standard finite difference method of sufficiently high order? The reason is that, in general, simple PDE solvers employing high-order centered difference methods in the interior of the domain and equally high-order biased stencils near the boundary are not stable. In addition, \cite{Lyon2009} demonstrate that the FC-Gram method exhibits significantly smaller pollution errors than finite difference and finite element methods.

\revtext{Note that one could use, and we have successfully tested, other sets of basis functions instead of Gram polynomials, depending on the problem at hand. While Gram polynomials are a natural choice due to their orthogonality on a uniformly discretized bounded domain, other function families may be preferable for different domains, grids, or boundary conditions.}

\revtext{In terms of parameter choices, we follow the recommendations in \cite{Lyon2009} and choose the extension size $N_{ext}$ such that the total computational domain has a simple prime factorization, as suggested in \citet{Albin2011}. Table \ref{tab:fcgramparams} details our default parameter choices.}

\begin{table}[ht]
    \centering
    \begin{tabular}{l l c}
        \toprule
        \textbf{Parameter} & \textbf{Meaning} & \textbf{Typical Value}\\
        $N$ & Physical wave function grid points & 16 \\
        $N_{ghost}$ & Size of ghost regions & 8\\
        $N_{\text{boundary}}$ & Boundary points for Gram expansion & 14\\
        $N_{ext}$ & Size of unphysical extension & 32\\
        $g$ & Fourier modes in SVD optimization & 63\\
        $\Lambda$ & Collocation points in SVD & 150\\
        $N_{\text{taylor}}$ & Taylor terms for kinetic operator & 7\\
        $\alpha$ & Exponential filter coefficient & $16 \log(10)$\\
        $\beta$ & Exponential filter exponent & $50$
    \end{tabular}
    \caption{\revtext{Typical parameters used in our FC-Gram implementation.}}
    \label{tab:fcgramparams}
\end{table}

Empirically, we find that the FC-Gram Schrödinger equation solver is stable if the time steps obey a CFL condition of the form
\begin{equation}
\label{eq:wave_cfl}
    \Delta t \leq C_G \frac{m}{\hbar}\Delta x^2,
\end{equation}
where $C_G = 0.2$. 
In the context of the FC-Gram algorithm, the allowed time step also depends on the size of the ghost region, as it is important to ensure that unphysical information from the extension domain does not propagate into the physical subdomain beyond the ghost region. While this cannot be entirely avoided due to the spectral nature of the algorithm, we observe that this choice of time step yields excellent results in practice. 

\subsubsection{Optimization with a Single Matrix Multiplication}
\label{subsec:SingleMatrix}
The asymptotic computational cost of this algorithm is dominated by the cost $\mathcal{O}(N \log N)$ for the Fast Fourier Transform (FFT). Determining the Gram polynomial coefficients requires two matrix-vector multiplications at a cost of $\mathcal{O}(N_{boundary}^2)$ operations. The cost of computing the extension is linear in the size of the unphysical domain. The time evolution operator also comes at a linear cost. 

\begin{figure*}[ht!]
\centering
\includegraphics[width=\textwidth, page=2]{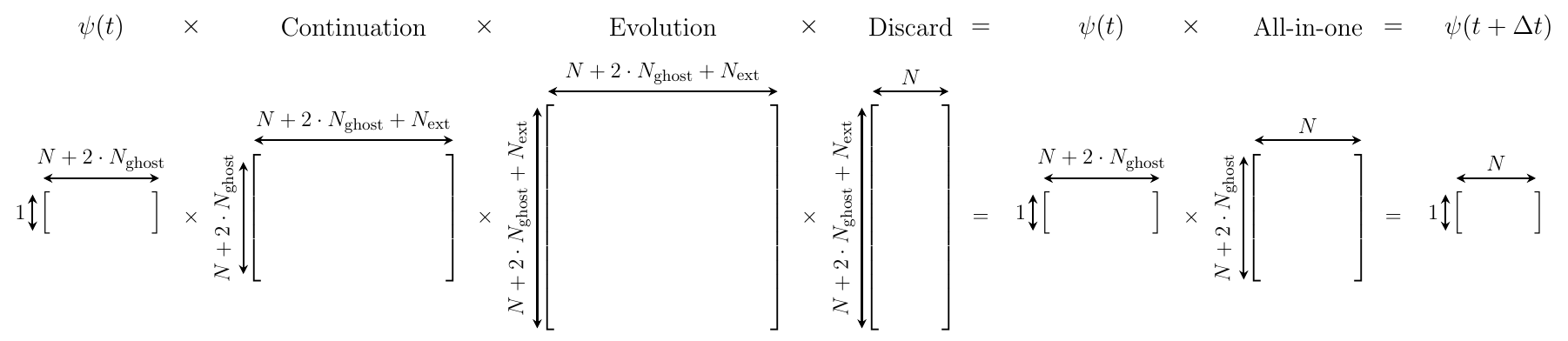}
\caption{Matrix representation of the FC-Gram solver for the free Schrödinger equation. The three matrix multiplications on the left-hand side of the equation can be combined into a single matrix multiplication with a smaller matrix. The combined matrix only needs to be precomputed once every time step since it depends only on the size of the time step and not on the wave function itself. This accelerates the FC-Gram Schrödinger equation solver for small domain sizes $N$.}
\label{fig:gramfe_matmul}
\end{figure*}

An additional improvement to the FC-Gram algorithm is that, for small domain sizes, it can be computationally efficient to combine the extension, time evolution operation, and discarding operations (second to fourth panels of Fig. \ref{fig:gramfe_fft}) into a single matrix operation. This is possible because both the FC-Gram algorithm and the solution of the kinetic operator in the Schrödinger equation, implemented via Fourier transforms, are linear operations that can be represented by a single linear map. The computational cost of this algorithm is given by a single matrix multiplication with $\mathcal{O}(N (N+2 N_{ghost}))$ operations. This simplification is shown in Figure \ref{fig:gramfe_matmul}.

\subsection{FC-Gram Interpolation}
\label{subsec:local_spectral_interpolation}
The FC-Gram algorithm is not only effective for building stable PDE solvers for wave equations but also facilitates the accurate interpolation of non-periodic data on a uniform grid. \revtext{In PDE simulations with adaptive mesh refinement, accurate interpolation is essential for reconstructing data on finer grids—whether during mesh refinement or when interpolating coarse-grid data to provide boundary conditions for a finer grid (see Section \ref{subsec:integration_into_gamer} for more details).} The input data is expanded in terms of Gram polynomials at the left and right boundaries. One then conducts a change-of-basis to replace the Gram polynomials with their precomputed periodic extensions. \revtext{The data can subsequently be interpolated without encountering the Gibbs phenomenon using a Fourier transform by directly evaluating the Fourier series.} Finally, the unphysical data are discarded. Although this interpolation method is neither monotonic nor conservative, it is highly accurate and represents an alternative to local polynomial interpolation. We choose a ghost region of $3$ points on either side of the input data. Interpolating $N$ points to $N_{int}$ interpolation points involves an asymptotic computational cost of $\mathcal{O}(N \log N)$ + $\mathcal{O}(N_{int} \log N_{int})$ operations. Furthermore, as discussed in Section \ref{subsec:SingleMatrix}, it can prove advantageous to combine the extension and interpolation operations into a single matrix multiplication with an asymptotic cost of $\mathcal{O}(N N_{int})$ operations.

\begin{figure*}[ht!]
\centering
\includegraphics[width=\textwidth]{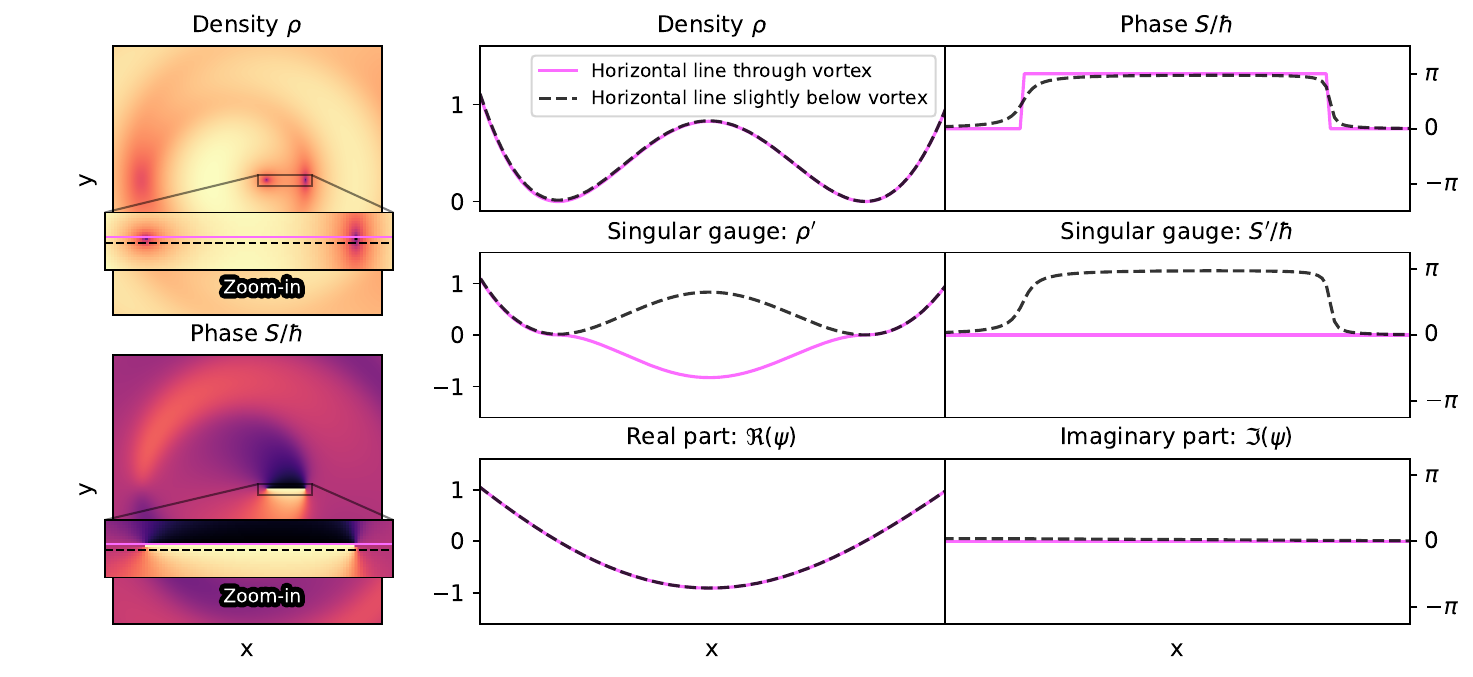}
\caption{Different representations of the wave function around a vortex pair described by Eq. \eqref{eq:VortexPairRotating}. Left: density (top) and phase (bottom) slices. Zoom-in plots highlight the horizontal one-dimensional density and phase lines through (solid lines) and close (dashed lines) to the vortices shown on the right. Right:  Density and phase (top), density and phase in singular gauge (middle), and real and imaginary parts (bottom) along the horizontal lines depicted in the left panels. While the density field at the vortex is smooth, the phase exhibits a jump by $\pi$ at the vortex and changes rapidly near the vortex. The singular gauge remedies the phase discontinuity at the vortex, but does not mitigate rapid changes of the phase field close to the vortex. In contrast, both the real and imaginary parts of the wave function are well-behaved around the vortex.}
\label{fig:singular_gauge}
\end{figure*}

In addition, we propose a new interpolation strategy: When interpolating the wave function in regions where the density is smooth and the velocity is high, it can be advantageous to interpolate the density $\rho$ and the phase $S$ instead of the wave function $\psi$. \revtext{For example, a plane wave with speed $v$ has a constant density and a phase $S = mvx$, which varies linearly with the spatial coordinate $x$. In contrast, its real and imaginary parts oscillate with wavelength $\lambda = h/(mv)$. In this case, interpolating the real and imaginary parts would require a resolution of at least two points per wavelength, even in an ideal setting using a Fourier transform with periodic boundary conditions. Conversely, a simple linear interpolation would be sufficient to accurately interpolate the density and phase. However, there is a significant drawback to this approach: It fails in regions of vanishing density where the phase suffers from a discontinuity of $\pi$.} This limitation is illustrated in Figure \ref{fig:singular_gauge}, which highlights the rotating vortex pair described by Eq. \eqref{eq:VortexPairRotating} on a grid with $N=512^2$ points at $t=0$, with all other parameters unchanged from Fig. \ref{fig:boundary_matching}. Consequently, a pseudospectral interpolation algorithm applied to $\rho$ and $S$ would introduce errors of order unity. We therefore implement a row-wise check for the continuity of the phase field. If the dimensionless second derivative of the phase field, discretized as $|S_{i+1} - 2 S_i + S_{i-1}|\reveq{/\hbar}$, exceeds a threshold, which we set to $0.1$, we interpolate the wave function $\psi$ instead of $\rho$ and $S$. Alternatively, one might consider the singular gauge transformation $\rho' = -\rho$ and $S' = S + \pi\reveq{\hbar}$. Although this leads to a smooth wave function \emph{at} the vortex, the phase still changes rapidly \emph{near} the vortex, as shown in Figure \ref{fig:singular_gauge}. We therefore prefer to directly interpolate the wave function around regions of vanishing density. 

\subsection{Madelung Refinement Criterion}
\label{subsec:madelung_refinement}
To ensure proper switching between the fluid and wave formulations of the Schrödinger equation, it is crucial to detect when and where the Hamilton-Jacobi-Madelung equations fail. This occurs at \revtext{regions with vanishing density}, where the quantum pressure diverges and the phase is discontinuous. We take advantage of this knowledge to design dimensionless refinement criteria based on the expressions $(\Delta x)^2 {\nabla}^2\sqrt{\rho}/\sqrt{\rho}$ and $(\Delta x)^2 {\nabla}^2 S\reveq{/\hbar}$, which determine when to increase resolution and switch from fluid to wave formulation. Specifically, we opt for a one-dimensional, second-order discretization of the criteria:
\begin{align}
\label{eq:flu_refine_threshold_c1}
    \mathcal{C}_{1} &=\left|\sqrt{\rho}_{i+1} - 2 \sqrt{\rho}_{i} + \sqrt{\rho}_{i-1}\right|/\sqrt{\rho}_{i},\\
\label{eq:flu_refine_threshold_c2}
    \mathcal{C}_{2} &= \left|S_{i+1} - 2 S_{i} + S_{i-1}\right|\reveq{/\hbar}.
\end{align}
If $\mathcal{C}_{1}$ or $\mathcal{C}_{2}$ exceeds a threshold set for each resolution level at runtime, the algorithm refines the region. In addition, it will switch to the wave scheme upon reaching a predetermined fluid-to-wave transition level. This transition level must be determined through trial and error as explained in Section \ref{subsec:integration_into_gamer}. We adopt the refinement criteria $\mathcal{C}_1 \geq 0.03$ and $\mathcal{C}_{2} \geq 1.0$ by default.

\begin{figure*}[ht!]
\centering
\includegraphics[width=\textwidth]{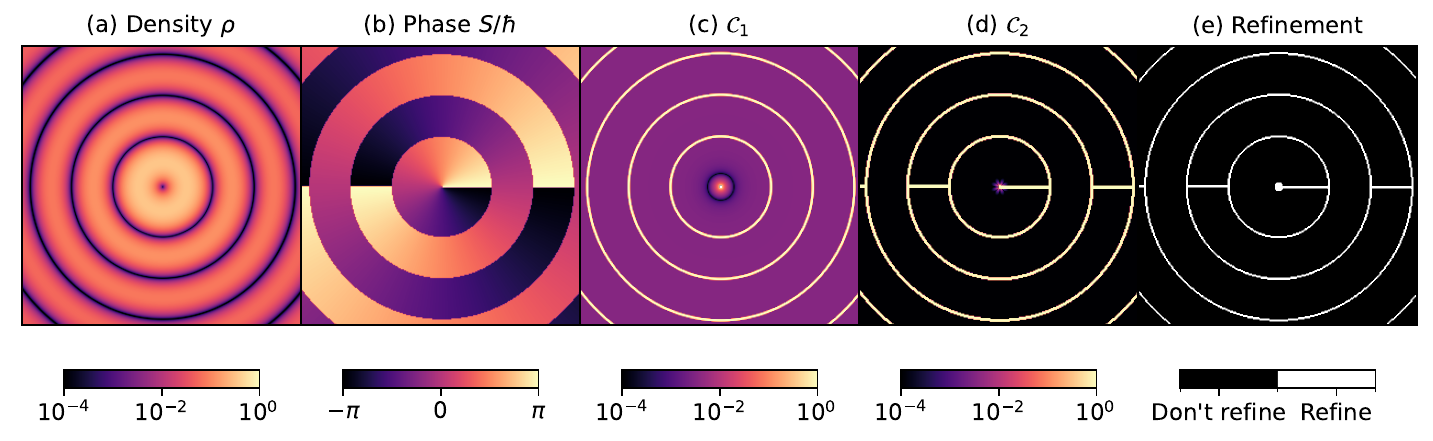}
\caption{Illustration of the Madelung refinement criterion Eqs. \eqref{eq:flu_refine_threshold_c1} and \eqref{eq:flu_refine_threshold_c2} in vortex rings described by Eq. \eqref{eq:VortexPairRotating}. The combined criteria reliably flag regions with vanishing density and discontinuous phase. From left to right: Density field, phase field, quantum pressure criterion $\mathcal{C}_{1}$, phase curvature criterion $\mathcal{C}_{2}$, and the refinement decision based on the thresholds $\mathcal{C}_{1}=0.03$ and $\mathcal{C}_{2}=1.0$.}
\label{fig:madelung_refinement}
\end{figure*}

The refinement process is illustrated in Fig. \ref{fig:madelung_refinement}, where the Madelung refinement criterion is applied to Eq. \eqref{eq:VortexPairRotating} with $\rho_{bg} = 0$, $A=1$, $\omega = 90$, $m/\hbar=1$, and $t=0$ in the two-dimensional domain $L = [-0.8, 0.8]^2$ with $N=256^2$ grid points. We refine if the quantum pressure criterion $\mathcal{C}_1 \geq 0.03$ or if the phase curvature criterion $\mathcal{C}_{2} \geq 1.0$. These thresholds are found to reliably flag simulation regions where destructive interference develops and have been verified in a wide range of one-, two-, and three-dimensional test problems, such as standing and traveling Gaussian waves, rotating and traveling vortex pairs, and cosmological simulations with CDM and FDM initial conditions.

These two refinement criteria serve a threefold purpose: Firstly, they ensure that the mesh resolution for the Hamilton-Jacobi-Madelung equations is increased when the density and phase fields develop small-scale features that would affect the numerical accuracy. Secondly, they warrant switching to the wave formulation of the Schrödinger equation before strong destructive interference sets in. Thirdly, $\mathcal{C}_{2}$ ensures that parts of the simulation domain containing phase discontinuities remain refined, even when the density field is smooth and non-vanishing. Why is this important?  This kind of phase discontinuity is a regular occurrence in a hybrid simulation using both the fluid and the wave solvers. The evolution of the Schrödinger equation using the wave solver leads to the formation of vortices, where pairs of vortices are connected by a phase jump of $2 \pi$ as shown in Figs. \ref{fig:boundary_matching} and \ref{fig:singular_gauge}. The fluid scheme cannot uniquely determine whether these phase jumps are physical, so it treats them as physical. Therefore, parts of the simulation domain between two vortices also require the usage of the wave solver. The phase refinement criterion ensures that such regions are properly refined. 

\subsection{Spectral Refinement Criterion}
\label{subsec:spectral_refinement}
Once the hybrid scheme switches to the wave formulation of the Schrödinger equation, ensuring that the de Broglie wavelength is properly resolved by the grid becomes of paramount importance. We suggest using a refinement scheme based on the polynomial expansion underlying the FC-Gram algorithm. The accuracy of this polynomial expansion can be assessed by examining the decay rate of the polynomial coefficients. If the coefficients decay quickly, it implies that the higher-order terms of the polynomial contribute significantly less to the overall value, and truncating the expansion series after a few terms does not discard much information. Additionally, this property directly translates into the accuracy of the FC-Gram algorithm: The more important the higher-order terms in the polynomial expansion, the less accurate the periodic extension becomes.

\begin{figure*}[ht!]
\centering
\includegraphics[width=\textwidth]{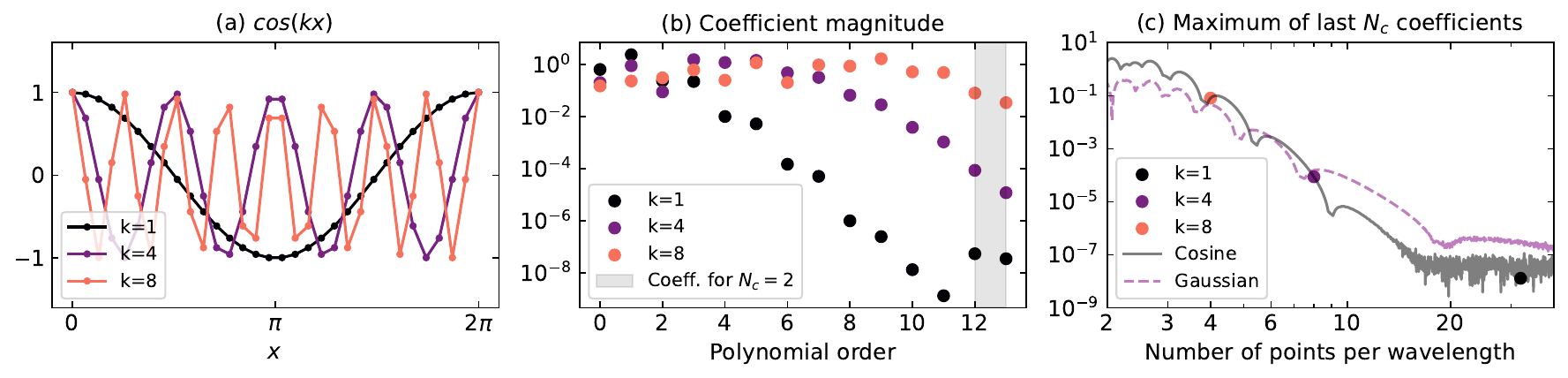}
\caption{Demonstration of the spectral refinement criterion. The plot shows that the magnitude of the coefficients of the polynomial expansion can serve as a refinement criterion. Input data that are well-resolved by the grid have small high-order coefficients. From left to right: (a) Three test functions with different wavenumbers $k$. (b) Their Gram polynomial expansions up to order $13$. (c) The maximum magnitude of the last $N_c = 2$ coefficients for the cosine function (solid line) as well as for the traveling Gaussian wave packet described by Eq. \eqref{eq:TravellingGaussian} (dashed line). The colored circles highlight the location of the cosine functions with $k=1$, $k=4$, and $k=8$. The input data are given in single precision.}
\label{fig:spectral_refinement}
\end{figure*}

In practice, we estimate the decay rate of the polynomial coefficients by checking the magnitude of the last $N_{c}$ coefficients. The grid is refined when at least one of these coefficients exceeds a given threshold $\mathcal{C}_{3}$. Fig. \ref{fig:spectral_refinement} showcases this refinement strategy for input functions with $N = 32$ points and an expansion in terms of polynomials up to order $13$. The plot indicates that a refinement threshold of $\mathcal{C}_{3} \sim 10^{-3}$ for the last $N_{c} = 2$ coefficients may be suitable for achieving around $6$ points per wavelength. While our tests indicate that this is true for simulations without gravity, we find that cosmological simulations with gravity and AMR require higher thresholds to avoid excessive refinement. In practice, we choose $N_{c} = 2$ coefficients and a threshold of $\mathcal{C}_{3}=1$ by default. Note that increasing $N_{c}$ makes the criterion more sensitive to higher resolutions, i.e. more than $20$ points per wavelength. 

\subsection{Integration into GAMER}  \label{subsec:integration_into_gamer}
The hybrid scheme and all its building blocks are implemented in the astrophysical AMR code \texttt{GAMER} and are parallelized using \texttt{MPI}, \texttt{OpenMP}, and \texttt{CUDA}. Specifically, the CPU version of the FC-Gram FFT algorithm is implemented using \texttt{FFTW2} and \texttt{FFTW3} \citep{Frigo2005}, and the GPU version utilizes Nvidia's \texttt{cuFFTDx}. The CPU version of the FC-Gram Matrix algorithm utilizes \texttt{GSL}'s BLAS routines \citep{GSL}, while the GPU version is custom-built in CUDA. Refinement and interpolation algorithms are CPU-only, with the FC-Gram interpolation algorithm implemented using \texttt{GSL}'s BLAS routines for the matrix version and \texttt{FFTW2}/\texttt{FFTW3} for the FFT version.

\subsubsection{Adaptive Mesh Refinement}
The hybrid scheme is integrated into \texttt{GAMER}'s AMR strategy, which subdivides the simulation domain into equal-sized, cubic subdomains called patches. These patches are organized using an octree, with varying refinement levels corresponding to the depth of each patch. The patches at the root node have the lowest resolution. Upon refinement, a patch is divided into eight children, each with twice the spatial resolution of their parent. Data exchange between refinement levels occurs bidirectionally: from parent to child patches via interpolation and from child to parent patches through averaging, ensuring that the average of eight high-resolution cells matches the value of the corresponding parent cell.

In implementing the hybrid algorithm, we employ either the fluid solver or the wave solver uniformly across all patches on the same refinement level, rather than on a per-patch basis. Although it might be beneficial in some scenarios to allow a mix of fluid and wave patches on the same refinement level, this approach is rarely required in cosmological FDM simulations, leading us to opt for a more straightforward implementation. Typically, the fluid scheme is used on the root level and several subsequent refinement levels. The first wave level is set as a user-specified parameter rather than being dynamically determined during runtime based on the mesh's ability to properly resolve the de Broglie wavelength. Because infall velocities increase over time in cosmological simulations, a grid sufficient to resolve the de Broglie wavelength at early times may become inadequate at later stages. Thus, the first wave level is established through trial and error, generally determined after one or two low-resolution test runs. If the first wave level is set too high, regions with destructive interference might be overly refined, resulting in unnecessary computational cost. Conversely, setting it too low may lead to numerical artifacts if the de Broglie wavelength at the fluid-wave boundary is not properly resolved.

\begin{figure*}[ht!]
\centering
\includegraphics[width=\textwidth]{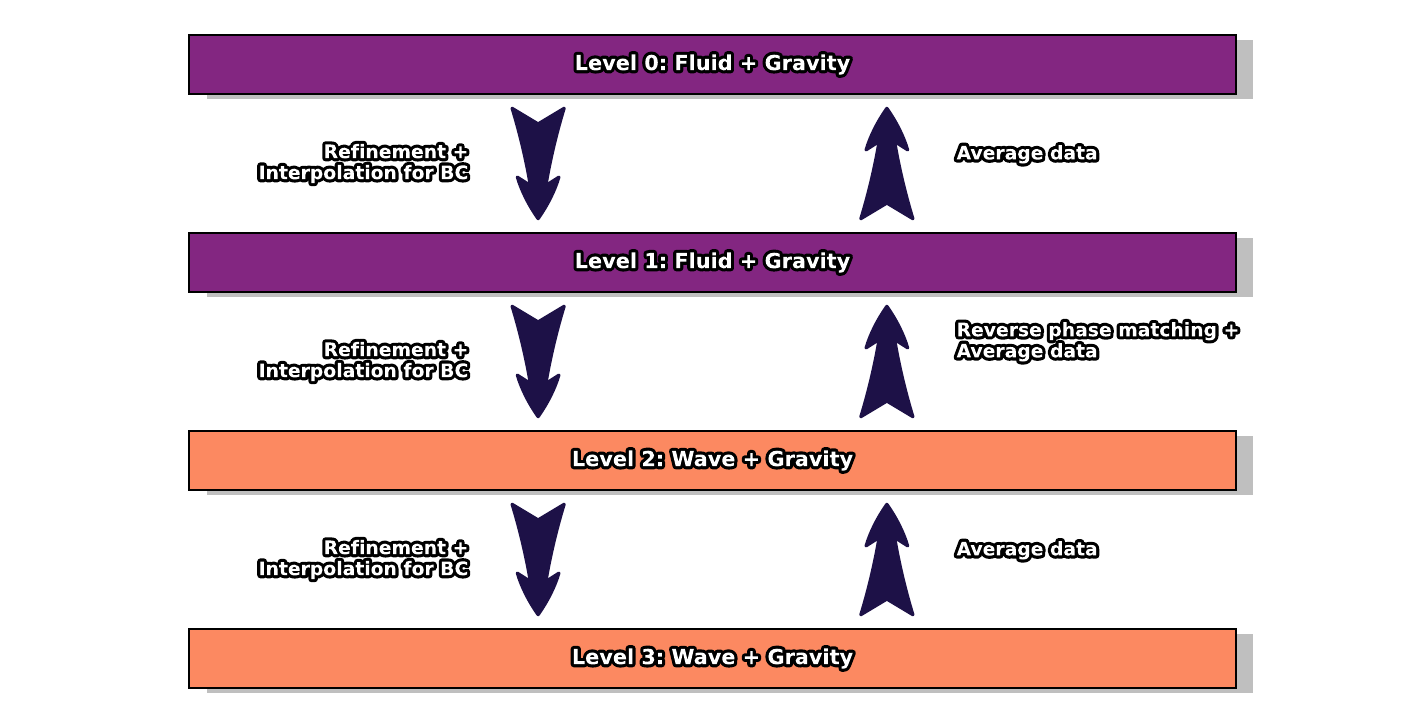}
\caption{Information flow for the hybrid scheme in \texttt{GAMER}, depicted using a simulation with two fluid and two wave levels. Information travels to higher-resolution levels through data interpolation, either during refinement or when setting boundary conditions for patches. Information on lower-resolution levels is updated by overwriting the fields with their averages from higher-resolution levels. Special attention is given to fluid-wave boundaries to properly handle the conversion from wave function to density and phase variables and vice versa (see Fig. \ref{fig:boundary_matching}).}
\label{fig:amr}
\end{figure*}

Fig. \ref{fig:amr} illustrates the global information flow of the hybrid scheme within \texttt{GAMER}, exemplified by a simulation using four resolution levels. The two coarser levels employ the fluid algorithm, while the finer two utilize the wave algorithm. Refinement for fluid-fluid and fluid-wave transitions is governed by Madelung refinement criteria (see Section \ref{subsec:madelung_refinement}). Wave-wave refinement can be influenced by a variety of criteria, including the spectral refinement criterion (see Section \ref{subsec:spectral_refinement}), energy and mass density-related criteria, as well as the Lohner error estimator. At the fluid-wave boundaries, the density and phase fields of the last fluid level are interpolated and then converted to the real and imaginary parts of the corresponding wave function. Conversely, when averaging data of the first wave level, the real and imaginary parts of the wave function are averaged and then converted back to the corresponding density and phase fields. The reverse boundary matching problem is addressed through unwrapping the average children cells' phase by adding a multiple of $2 \pi$ to match the parent cell's phase.

\texttt{GAMER} provides several runtime options to improve the interpolation and averaging steps: For the wave-wave level interpolation, users can choose whether to interpolate the real and imaginary parts of the wave function or the density and phase. 
When employing the new FC-Gram interpolation scheme and enabling density and phase interpolation, the interpolation algorithm automatically falls back to the interpolation of the real and imaginary parts close to vortices, as explained in Section \ref{subsec:local_spectral_interpolation}. \texttt{GAMER} also allows users to choose between averaging the density and phase fields or the real and imaginary parts of the wave function for the averaging operation. If the averaging and interpolation operations are consistent, their combined application, that is, first interpolating and then averaging, yields an identity map. If, however, one interpolates the density and phase fields and averages the real and imaginary parts, the nonlinear transformation between the two introduces a small reconstruction error.

In practice, \texttt{GAMER} always performs the averaging and interpolation operations on density for both the fluid-fluid and wave-fluid cases. For the wave-wave case, \texttt{GAMER} either averages/interpolates the density and phase directly or averages/interpolates the wave function first and then rescales its amplitude to match the averaged/interpolated density. It ensures consistent mass density between fluid-fluid, wave-fluid, and wave-wave levels, which is critical for the multi-level Poisson solver.

Mass can be conserved in the AMR implementation of the Hamilton-Jacobi-Madelung equations in the \texttt{GAMER} code. This is achieved by storing the mass fluxes $F_{i+\frac{1}{2}}$ defined in Eq. \eqref{eq:conservationlawfluxform} at the coarse-fine resolution boundaries and then correcting mismatches between the coarse- and fine-grid fluxes in a separate flux fix-up operation. 

\subsubsection{Adaptive Time Steps}
\texttt{GAMER} supports adaptive time steps to enhance computational efficiency. Using the same time step for all resolution levels usually results in the time step being determined by the highest resolution level, making it unnecessarily small on coarse levels. This can significantly impact overall performance, particularly for FDM simulations where time steps generally scale as $\Delta t \propto \Delta x^2$ (Eqs. \ref{eq:fluid_cfl} and \ref{eq:wave_cfl}). To address this, \texttt{GAMER} allows each resolution level to evolve using its own time step, where lower levels are advanced first and then wait until they are synchronized with higher levels.

This works as follows: We begin at time $t$ when all AMR levels are synchronized, and compute a time step $\Delta t_0$ on level $0$ that respects the CFL condition of the patches on level $0$. The solution on level $0$ is then advanced to time $t + \Delta t_0$ via a drift and a kick step. Next, we advance the solution on level $1$ to the time $t + \Delta t_1$, where $\Delta t_1 \leq \Delta t_0$ and $\Delta t_1$ respects the CFL condition of the patches on level $1$. The boundary conditions are provided by the patches on level $0$ through spatial interpolation of the solution at time $t$. We then proceed to higher levels until they are all synchronized with level $1$. Afterward, we interpolate the solution on level $0$ between $t$ and $t + \Delta t_0$ linearly in time to obtain an estimate of the solution on level $0$ at the time $t +\Delta t_1$. This estimate provides the boundary conditions for the second time step on level $1$ from $t + \Delta t_1$ to $t + \Delta t_1 + \Delta t_1'$, where $\Delta t_1 + \Delta t_1' \leq \Delta t_0$ and $\Delta t_1'$ respects the CFL condition for the patches on level $1$ at $t + \Delta t_1$. This process is repeated iteratively so that each AMR level evolves in time according to its own CFL condition while remaining synchronized with coarser levels.

This adaptive time-step algorithm is only first-order accurate in time and voids the theoretical convergence guarantees for the Schrödinger-Poisson system. However, in practice, it provides a significant speed-up by focusing computational resources on the higher resolution levels with more stringent time step requirements. Nevertheless, there is one major drawback to this approach: Since the solution on coarse levels with refined counterparts may be poorly resolved in the wave formulation or contain vortices in the fluid formulation, the time evolution on these levels may be very inaccurate or even unstable. \revtext{Still, evolving these cells remains necessary because they provide the boundary conditions for refined grids adjacent to coarse-fine resolution interfaces. For example, consider a two-level scenario where level $0$ uses a time step $\Delta t$ and level $1$ requires a time step $\Delta t/4$. The evolution of level $1$ over $\Delta t$ requires four substeps with corresponding boundary conditions at intermediate times.  Direct use of the refined wave function would assume frozen boundary conditions during these intervals, yielding a zeroth-order error in time.}

To mitigate this, special care is taken to ensure that the algorithm for the Hamilton-Jacobi-Madelung equations remains stable even when updating a solution with phase discontinuities and vortices. \revtext{This is crucial because cells with a refined counterpart are excluded from the CFL condition; otherwise, the high velocities in regions of vanishing density would force prohibitively small time steps.  The trade-off is that the maximum phase update $\Delta S = |S^{n+1} - S^n|$ can exceed $\pi\reveq{\hbar}$ and that negative densities can occur in fluid cells with a refined wave counterpart. To address this, we overwrite any negative density values with the time evolution computed using a second-order accurate finite-difference discretization of the Schrödinger equation, thereby avoiding the errors associated with frozen boundary conditions. In addition, our refinement criteria are meticulously adjusted to ensure that vortices occur at a safe distance from fluid-wave boundaries (see Section \ref{subsec:madelung_refinement}). In practice, we have observed that negative densities occur only in cells with a refined counterpart when the CFL condition is violated; when the CFL condition is met, the MUSCL-type algorithm for solving the continuity equation effectively prevents such instabilities.}

\section{Results} \label{sec:results}
In this section, we first present tests evaluating the accuracy of the numerical schemes for solving the Hamilton-Jacobi-Madelung and the Schrödinger equations presented in Sections \ref{subsec:hamilton_jacobi} and \ref{subsec:local_spectral_solver}. We also demonstrate the precision of the FC-Gram interpolation algorithm presented in Section \ref{subsec:local_spectral_interpolation}. Next, we compare the performance of different FDM algorithms implemented in \texttt{GAMER} and evaluate the strong scaling of the hybrid scheme. We then compare the Hamilton-Jacobi-Madelung solver with alternative approaches for modeling the evolution of the power spectrum on large scales in a cosmological FDM simulation within a $L=10$ Mpc/h box. Finally, we provide a detailed analysis of a cosmological FDM simulation within a $L=5.6$ Mpc/h box using the hybrid scheme, as previously illustrated in Fig. \ref{fig:madelung_schroedinger_mismatch}.

\subsection{Accuracy}
\subsubsection{Fluid and Wave Solvers}
\label{subsec:solver_accuracy}
\begin{figure*}[ht!]
\centering
\includegraphics[width=\textwidth]{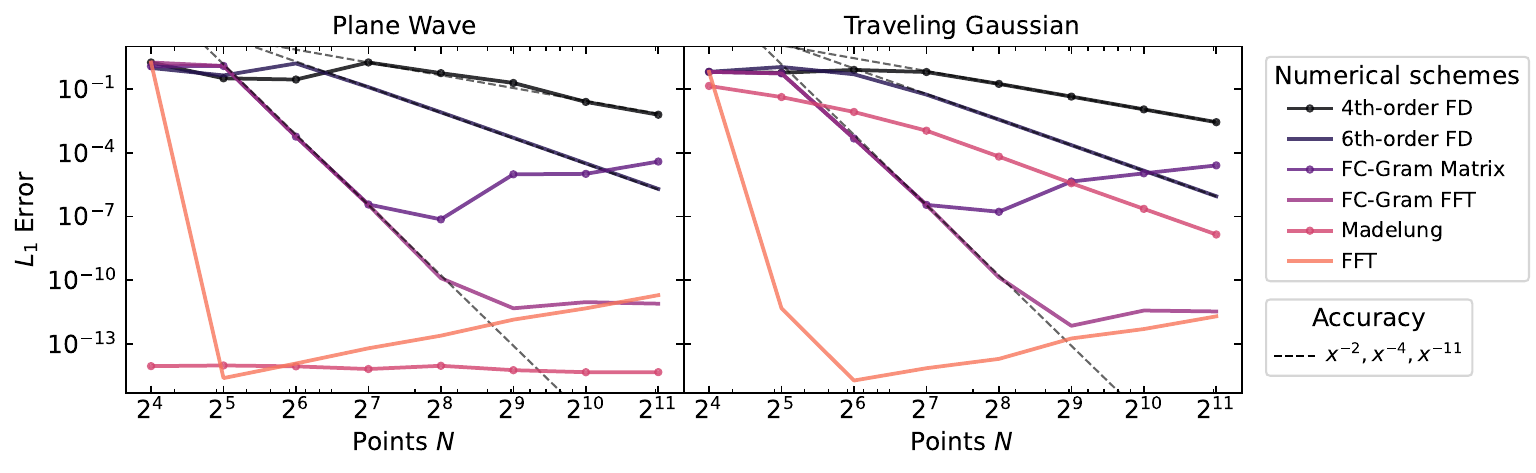}
\caption{Mean $L_1$ errors of the real and imaginary parts of the wave function for the plane wave and traveling Gaussian wave packet tests, as described by Eqs. \eqref{eq:PlaneWave} and \eqref{eq:TravellingGaussian}, plotted as a function of the number of grid points $N$. Here, $N = 2^5$ roughly corresponds to two points per wavelength. The FC-Gram algorithms have a higher resolution power and convergence rate compared to finite difference methods. Yet, they still fall short of the accuracy achieved by the root-level pseudospectral FFT solver (FFT). The Hamilton-Jacobi-Madelung solver (Madelung) exhibits a convergence rate similar to the finite difference methods in the traveling Gaussian wave packet test but, importantly, remains effective even for resolutions below two points per wavelength.}
\label{fig:error_vs_wavelength}
\end{figure*}

Fig. \ref{fig:error_vs_wavelength} compares the accuracy of various FDM algorithms implemented in \texttt{GAMER} for different resolutions using two one-dimensional test problems: a plane wave 
\begin{equation}
    \label{eq:PlaneWave}
    \psi = \exp(ikx),
\end{equation}
with $k=32 \pi$ and $\hbar/m=1$ on a domain of size $L=1$ evolved from $t=0$ until $t=0.015$ as well as a moving Gaussian wave packet described by 
\begin{equation}
    \label{eq:TravellingGaussian}
    \psi(x,t) = \sqrt{\frac{\alpha}{\alpha + i t \frac{\hbar}{m}}} \exp\left(-\frac{(x - x_0 - i k \alpha)^2}{2(\alpha + i t \frac{\hbar}{m})} -\frac{\alpha k^2}{2}\right),
\end{equation}
for $\alpha = 0.04$, $\hbar/m=1/64$, $k = 32 \pi$, and $x_0 = 0.3$ on a domain of size $L=1$ evolved from $t=0$ until $t=0.318$. Time steps are chosen according to the default CFL condition of each algorithm (see Eqs. \eqref{eq:fluid_cfl} and \eqref{eq:wave_cfl}). Both test problems use periodic boundary conditions except for the Madelung algorithm tests. In the case of the Gaussian wave packet, the periodic boundary conditions are enforced by adding periodic images to the initial condition at both sides of the simulation domain. The Madelung algorithm tests use boundary conditions provided by the analytical solutions of the test problems. This is necessary because the phase field of the plane wave is not periodic, and the phase field of the periodic Gaussian wave packet, defined via an inverse tangent, is generally discontinuous.

The test problem parameters are chosen such that $N = 2^5$ points roughly correspond to two points per wavelength. For the plane wave, this relation is exact. For the Gaussian wave packet, the relation is approximate, as the number of points per wavelength is given as $N (2\pi/k)$ at $t = 0$ but becomes position-dependent at later times, partly due to self-interference caused by the periodic boundary conditions. All accuracy tests are carried out in double-precision floating-point arithmetic.

Figure \ref{fig:error_vs_wavelength} shows the results of the accuracy test. The FC-Gram algorithms exhibit $11$th-order convergence starting at roughly two points per wavelength, significantly outperforming the $4$th-order and $6$th-order finite difference methods that respectively exhibit $2$nd-order convergence starting at approximately $8$ points per wavelength and $4$th-order convergence starting at approximately $4$ points per wavelength. The reported convergence orders reflect the spatial accuracies of the adopted algorithms minus two because of the need to increase the number of time steps quadratically with the spatial resolution. 
We conclude that the spatial accuracy of the FC-Gram algorithms is $13$th-order, which is one order less than expected from the discussion in Section \ref{subsec:local_spectral_solver}. This discrepancy is due to the numerical ill-conditioning of high-order polynomial expansions.

The FC-Gram FFT and Matrix algorithms perform similarly for low resolutions. For high resolutions, the FFT algorithm converges to machine precision and then exhibits an $\propto N^2$ increase in error because of the need to adopt ever smaller time steps for higher resolutions. In contrast, the FC-Gram Matrix algorithm has a higher $L_1$ error floor of roughly $10^{-7}$ despite both tests using double-precision floating-point arithmetic. The reason why the FC-Gram Matrix algorithm does not converge to double-precision machine precision is that the matrix representing the FC-Gram Matrix algorithm is ill-conditioned. Therefore, we recommend the FFT algorithm for simulations using double-precision floating-point arithmetic.

Yet, the matrix algorithm offers one notable advantage over the FFT algorithm: The latter requires double-precision floating-point arithmetic for stability according to our tests. In contrast, the matrix multiplication algorithm remains stable even when carried out in single-precision floating-point arithmetic, in which case the $L_1$ error has a floor of roughly $10^{-6}$, only slightly higher than in the double-precision case. Additionally, single precision is substantially faster on desktop GPUs.

From Fig. \ref{fig:error_vs_wavelength}, it is clear that none of the wave algorithms is competitive with a pseudospectral method using an FFT for periodic boundary conditions, whose convergence starts at two points per wavelength. At even lower resolutions, we observe that the Hamilton-Jacobi-Madelung solver remains effective for less than two points per wavelength, i.e., when the phase difference between neighboring points exceeds $\pi$.

\begin{figure*}[ht!]
\centering
\includegraphics[width=\textwidth]{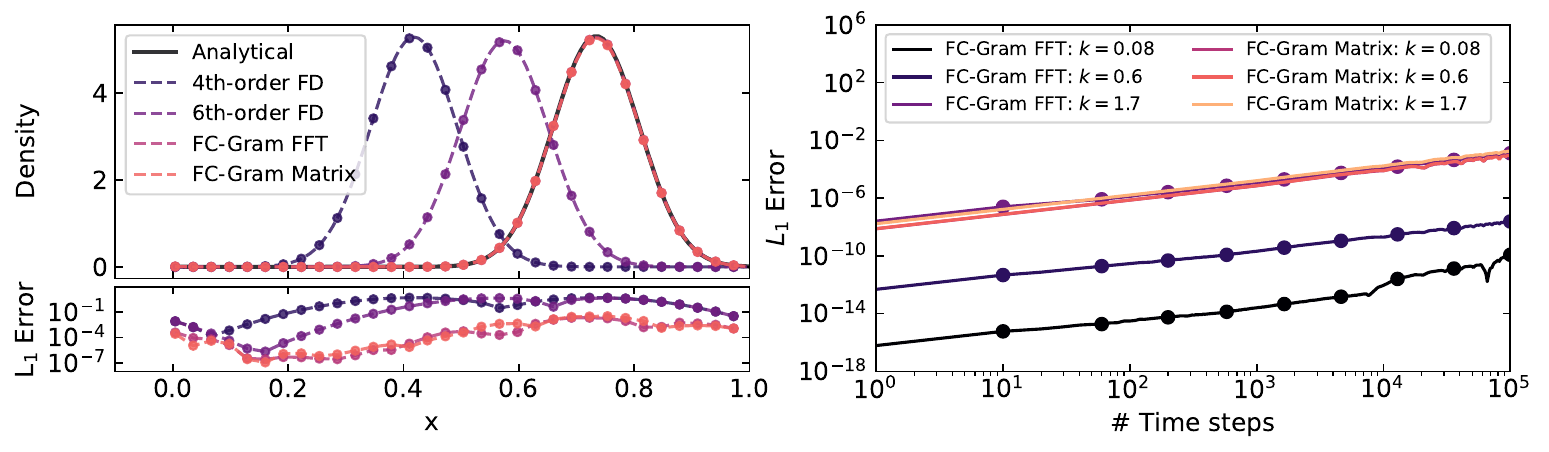}
\caption{Accuracy and stability tests with traveling Gaussian wave packets according to Eq. \eqref{eq:TravellingGaussian}. The finite difference solutions lag behind the analytical solution, whereas the FC-Gram schemes demonstrate noticeably higher accuracy (left). Moreover, the FC-Gram schemes remain stable over $10^5$ time steps (right). The FC-Gram Matrix scheme uses single-precision floating-point arithmetic while the FC-Gram FFT scheme uses double-precision floating-point arithmetic in these tests. Despite this difference, the two schemes agree well for wave functions with high velocities.}
\label{fig:gaussian_lag}
\end{figure*}

Fig. \ref{fig:gaussian_lag} compares the densities of the finite difference and the FC-Gram algorithms (left panel) for the traveling Gaussian wave packet test described by Eq. \eqref{eq:TravellingGaussian} and demonstrates the long-term stability of the FC-Gram algorithm (right panel). The parameter choices are $N=128$ points, $L=1$, $\hbar/m=1/70$, $\alpha=0.01$, and $k=210$ for the solver comparison at the final time $t=0.26$ after $513$ time steps, as well as $k\sim0.077$, $0.599$, and $1.668$ for the FC-Gram stability test. We observe a significant advantage of the FC-Gram algorithms: They resolve velocities more accurately than finite difference schemes at the same resolution. The solutions evolved using the finite difference algorithms noticeably lag behind the analytical solution, even though mass is still conserved by also solving the continuity equation in \texttt{GAMER}.

The long-term stability test indicates that numerical errors grow linearly with the number of time steps. At higher resolutions ($k\sim0.077$), the error pattern seems to become irregular after around $10^4$ steps. However, in practice, we have not observed any instabilities with the recommended parameter settings. We have also extensively tested cosmological simulations in single-precision floating-point arithmetic using the FC-Gram Matrix algorithm and the results are stable.

\subsubsection{Interpolation}

\begin{figure*}[ht!]
\centering
\includegraphics[width=\textwidth]{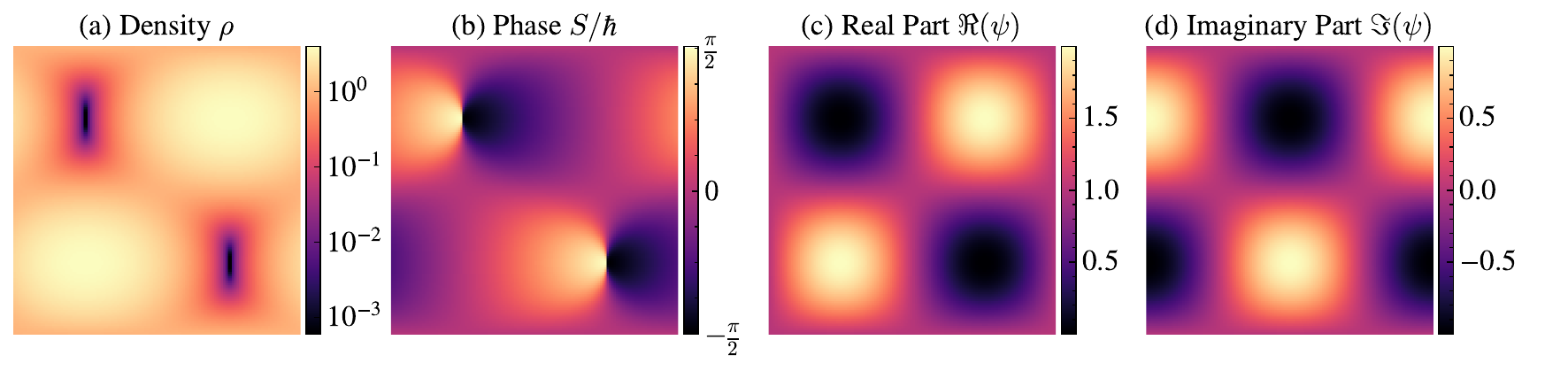}
\caption{Traveling vortex pair test described by Eq. \eqref{eq:LinearVortexPair}. At the vortices, the density field (a) approaches zero and the phase field (b) is discontinuous, whereas both the real (c) and imaginary parts (d) of the wave function remain well-behaved. The vortices travel periodically in the positive x-direction.}
\label{fig:vortex_pair_linear}
\end{figure*}

This section compares the interpolation errors of the conservative and monotonic polynomial interpolation schemes implemented in \texttt{GAMER} with the FC-Gram Matrix interpolation algorithm (see Section \ref{subsec:local_spectral_interpolation}). It showcases two test problems: A one-dimensional plane wave with $\hbar/m = 1$ and varying $k_x$ in the domain $L = [0, 1]$ with $N=512$ points, and a two-dimensional traveling vortex pair described by 
\begin{equation}
\label{eq:LinearVortexPair}
    \psi(x,y,t) = \rho_{bg} + A \cos(k_y y) \exp\left(i \left( k_x x - \omega t/2 \right)\right),
\end{equation}
with the parameters $\rho_{bg} = 1$, $A=1$, $k_y = 1$, $\omega = (\hbar/2m)(k_x^2 + k_y^2)$, $\hbar/m = 1$, and varying $k_x$ in the domain $L = [0.25, 1.25]^2$ with $N=64^2$ points. The traveling vortex pair test is depicted in Fig. \ref{fig:vortex_pair_linear}.
The input data are divided into grid patches of $8^2$ points with a ghost region of size $2$ on all sides. Therefore, the input data seen by the interpolation algorithms have $12$ points per dimension. The wavelength is calculated using $\lambda = 2\pi/k_x$. For the traveling vortex pair, only $k_x$ is varied, while $k_y$ is kept constant.

\begin{figure*}[ht!]
\centering
\includegraphics[width=\textwidth]{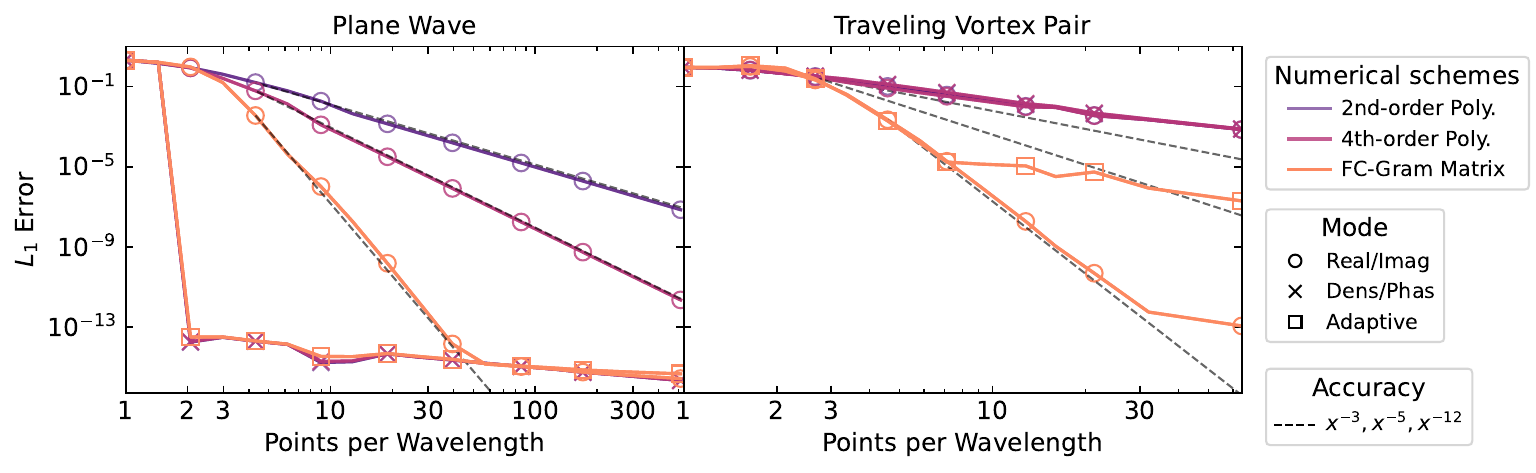}
\caption{Interpolation errors for the plane wave (left) and traveling vortex pair (right) described by Eqs. \eqref{eq:PlaneWave} and \eqref{eq:LinearVortexPair}, respectively. The FC-Gram Matrix interpolation method converges faster than the 2nd-order and 4th-order polynomial interpolation methods. The mode determines whether the real and imaginary parts or the density and phase are interpolated for polynomial interpolation. If the solution does not contain vortices, such as in the plane wave test, interpolating the density and phase fields gives more accurate results than interpolating the wave function. In the adaptive mode used for the FC-Gram Matrix interpolation, the density and phase are interpolated by default, while the wave function is interpolated around vortices.}
\label{fig:interpolation_error}
\end{figure*}

Fig. \ref{fig:interpolation_error} shows the interpolation results. For the plane wave test, we observe the expected convergence behavior when interpolating the real and imaginary parts of the wave function. Polynomial interpolation of order $n$ yields an $(n+1)$th-order accurate method. The FC-Gram Matrix interpolation algorithm is approximately $12$th-order accurate. This is because an extension domain containing $N_{boundary}$ points yields an extension with $N_{boundary}-1$ continuous derivatives and therefore an $N_{boundary}$th-order accurate method, as explained in Section \ref{subsec:local_spectral_solver}. When interpolating the density and phase instead, we observe that all three schemes exhibit the same convergence, as both the density and phase of the plane wave are linear functions \revtext{of space}.

For the traveling vortex pair test, the difference between the interpolation methods becomes more pronounced. When interpolating the real and imaginary parts, the FC-Gram interpolation algorithm exhibits the expected convergence behavior, while the polynomial interpolations reduce to first-order convergence. This reduction occurs because the monotonicity requirement introduces errors for this more challenging test problem. Likewise, convergence of the polynomial interpolations is slow when interpolating the density and phase fields. This is because the phase is discontinuous at the vortex, as highlighted in Figs. \ref{fig:singular_gauge} and \ref{fig:vortex_pair_linear}. The FC-Gram Matrix interpolation scheme avoids this problem by interpolating the density and phase by default but switching to interpolating the real and imaginary parts around vortices (see Section \ref{subsec:local_spectral_interpolation}). However, the convergence of the interpolation error still stalls at around seven points per wavelength. This is because most of the interpolation domain is interpolated using the density and phase fields.

\subsection{Performance}
\begin{figure*}[ht!]
\centering
\includegraphics[width=\textwidth]{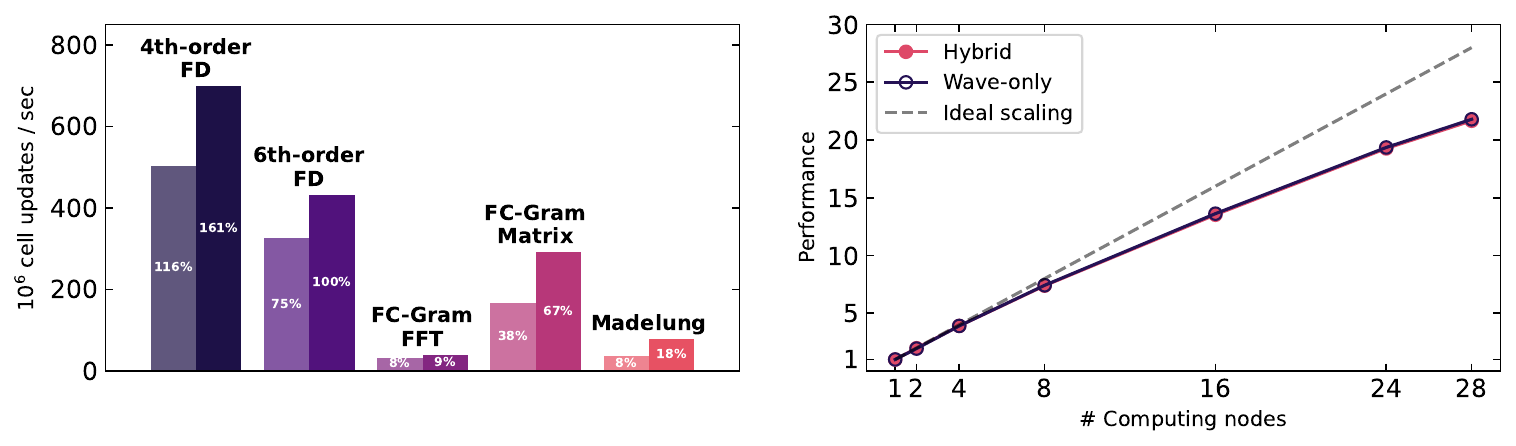}
\caption{Left panel: Single-node performance of various FDM algorithms, where the light and dark colors in every pair of bars represent CPU and GPU performance, respectively. Right panel: Strong scaling test on up to $28$ computing nodes using \texttt{OpenMP}, \texttt{MPI}, and GPUs. The FC-Gram Matrix algorithm (FC-Gram Matrix) exhibits a $75\%$ speed-up from running on a GPU and is only $33\%$ slower than the default 6th-order finite difference method adopted in \texttt{GAMER} ($6$th-order FD). When scaled to $28$ computing nodes, both the hybrid and wave-only schemes using the FC-Gram Matrix algorithm achieve a speed-up of $21.8$ compared to a single node. Each node uses four \texttt{MPI} processes, eight \texttt{OpenMP} threads per process, and one GPU. All results in this plot, except those from the FC-Gram FFT solver, were computed using single-precision floating-point arithmetic.}
\label{fig:performance}
\end{figure*}

Fig. \ref{fig:performance} compares the single-node performance of various FDM algorithms implemented in \texttt{GAMER} and showcases the strong scaling of the hybrid scheme on up to 28 computing nodes. Each node is equipped with an AMD Ryzen Threadripper PRO 5975WX 32-Cores CPU, an NVIDIA GeForce RTX 3080 Ti GPU, and 256 GB of RAM. The performance tests are carried out with single-precision floating-point arithmetic (except for the internal calculations in the FC-Gram FFT solver), which is the default configuration for cosmological FDM simulations in \texttt{GAMER}.

We observe that the finite difference algorithms outperform both the FC-Gram solvers and the Hamilton-Jacobi-Madelung solvers because of their lower number of floating-point operations. The FC-Gram FFT solver exhibits only marginal speed-up from using a GPU (dark color) compared to its CPU implementation (light color). This limited GPU acceleration is partially because the FC-Gram FFT algorithm requires the use of double-precision operations (double-precision FFT and matrix multiplication), as described in Section \ref{subsec:solver_accuracy}, and the performance is measured on a desktop GPU with relatively low double-precision performance. In contrast, the FC-Gram Matrix algorithm, running in single-precision floating-point arithmetic using a GPU, exhibits 75\% and 600\% speed-ups over its CPU counterpart and the GPU FFT solver, respectively. The Hamilton-Jacobi-Madelung solver exhibits a 100\% speed-up through GPU acceleration.

The strong scaling of the hybrid scheme is comparable to that of the wave-only solver implemented in \texttt{GAMER} for the FC-Gram Matrix algorithm running on GPU. It achieves approximately $78\%$ parallel efficiency when scaling to $28$ nodes, with a total of $112$ MPI processes and $8$ OpenMP threads per process. In theory, the strong scaling of the hybrid scheme could be impacted by the data communication required to identify fluid cells with wave counterparts on refined levels. This information is used in two places. Firstly, it is used to determine the correct velocity for the Hamilton-Jacobi-Madelung velocity CFL condition (see Eq. \ref{eq:fluid_cfl}) since this condition is only computed for the fluid cells without wave counterparts on refined levels. Otherwise, phase discontinuities (see Fig. \ref{fig:madelung_refinement}) can lead to prohibitively small time steps. Secondly, it is used to decide when to correct the fluid solver with a second-order finite difference discretization of the wave equation around vortices (see Section \ref{subsec:integration_into_gamer}). In practice, however, the cost of this additional data communication is negligible.

\subsection{Linear Power Spectrum Evolution}
\begin{figure*}[ht!]
\centering
\includegraphics[width=\textwidth]{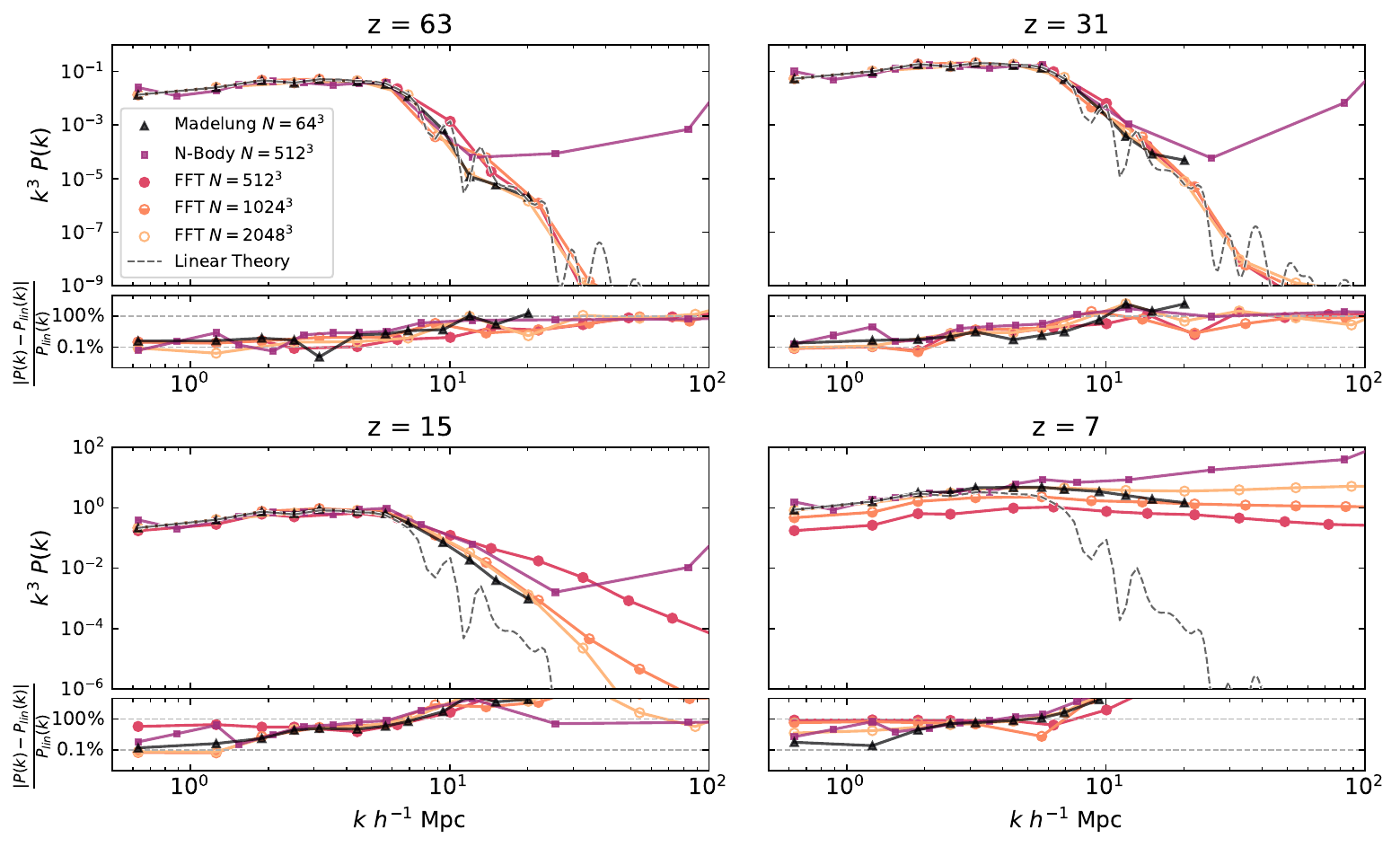}
\caption{Density power spectrum evolution of FDM cosmological simulations in a cubic comoving box with a side length of $10$ Mpc/h. The four panels show the dimensionless power spectra at different redshifts $z$ for three simulation methods: the Hamilton-Jacobi-Madelung solver, the collisionless N-body solver, and the root-level pseudospectral FFT solver. In addition, the results are compared with linear perturbation theory, $P_{lin}(k) = (a^2/a_0^2) P(k, a_0)$, where $a_0$ is the initial scale factor at $z=100$. The Hamilton-Jacobi-Madelung algorithm accurately tracks the linear evolution of the low-wave number modes even at $z=7$ with a grid of only $N=64^3$ points. In contrast, the root-level FFT algorithm requires more than $N=1024^3$ points to achieve similar accuracy.}
\label{fig:power_spectrum}
\end{figure*}

Fig.  \ref{fig:power_spectrum} shows the evolution of the density power spectrum of a cosmological simulation in a periodic $L=10$ Mpc/h box for the FDM mass $m = 8\times 10^{-23}$ eV. The initial power spectra at $z=100$ are created using \texttt{axionCAMB} and then converted into three-dimensional initial conditions using \texttt{MUSIC}. The N-body simulation was conducted using \texttt{GADGET-2} \citep{Springel2005}. The FDM simulations use uniform grids without refinement. The run using the Hamilton-Jacobi-Madelung solver utilizes $N=64^3$ grid points and does not use the hybrid scheme. The three root-level FFT runs use $N=512^3$, $1024^3$, and $2048^3$ points, respectively.

We observe that the Hamilton-Jacobi-Madelung algorithm accurately tracks the linear evolution of the low-wave number modes even at $z=7$ using only $N=64^3$ grid points. It shows good agreement with the N-body run using $N=512^3$ particles and the root-level FFT run with $N=2048^3$ grid points. In contrast, the lower-resolution root-level FFT runs underestimate the correlation on large scales. Finally, it is worth noting that the Hamilton-Jacobi-Madelung equations agree with the high-resolution FFT run in predicting a slight suppression of power compared to the N-body simulation in the weakly nonlinear regime for $5\,h\,{\rm{Mpc}}^{-1} \lesssim k \lesssim 20\,h\,{\rm{Mpc}}^{-1}$ at $z=7$. This suppression is expected from the effects of the quantum pressure.

\subsection{Hybrid Cosmological Simulation}
Finally, we study the cosmological simulation employing the hybrid scheme shown in Fig. \ref{fig:madelung_schroedinger_mismatch}. It features an FDM mass $m = 2\times 10^{-23}$ eV in a periodic simulation cube with a side length of $L=5.6$ Mpc/h. The initial conditions are generated using \texttt{axionCAMB} and \texttt{MUSIC} at $z=100$ on a uniform grid with $N=128^3$ points, referred to as level $0$. The fluid algorithm is used up to refinement level $3$, corresponding to an effective resolution of $N=1024^3$. The FC-Gram Matrix algorithm is used from levels $4$ to $7$, corresponding to effective resolutions of $N=2048^3$ to $16384^3$ grid points. The fluid-fluid and fluid-wave refinement is controlled by the Madelung refinement criterion, with thresholds of $C_1=0.03$ for the quantum pressure term and $C_2=1.0$ for the second derivative of the phase field. The wave-wave refinement is controlled by the spectral refinement criterion, with a coefficient threshold of $\mathcal{C}_{3}=1$ for the $N_c = 2$ last polynomial expansion coefficients. The simulation was carried out using $8$ computing nodes, each equipped with an AMD Ryzen Threadripper PRO 5975WX $32$-Cores CPU, an NVIDIA GeForce RTX 3080 Ti GPU, and $256$ GB of RAM. It took less than $4$ hours to reach $z=1$.

\begin{figure*}[ht!]
\centering
\includegraphics[width=\textwidth]{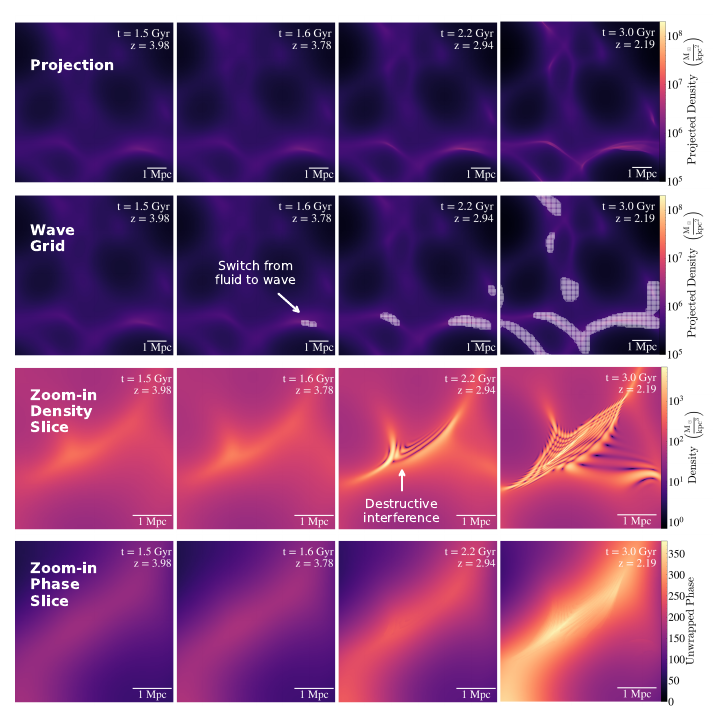}
\caption{Madelung refinement criterion in the cosmological simulation shown in Fig. \ref{fig:madelung_schroedinger_mismatch}. The wave solver is first employed at redshift $z=3.83$ before destructive interference first occurs at around $z=3.1$. Interference patterns are resolved using the wave scheme at later times. Left to right: Simulation snapshots at $z=3.98$, $3.78$, $2.94$, and $2.19$. Top to bottom: Density projection; density projection with white grids highlighting subdomains where the wave scheme is employed; zoom-in density slice through the bottom-right region of the projection plots where the wave scheme is first employed at $z=3.83$; zoom-in slice of the unwrapped phase field corresponding to the density slice.}
\label{fig:filament}
\end{figure*}

Fig. \ref{fig:filament} demonstrates the effectiveness of the Madelung refinement criterion. The wave solver is activated at redshift $z=3.83$ before destructive interference first occurs at around $z=3.1$. Interference patterns are resolved using the wave scheme at later times.

\begin{figure*}[ht!]
\centering
\includegraphics[width=\textwidth]{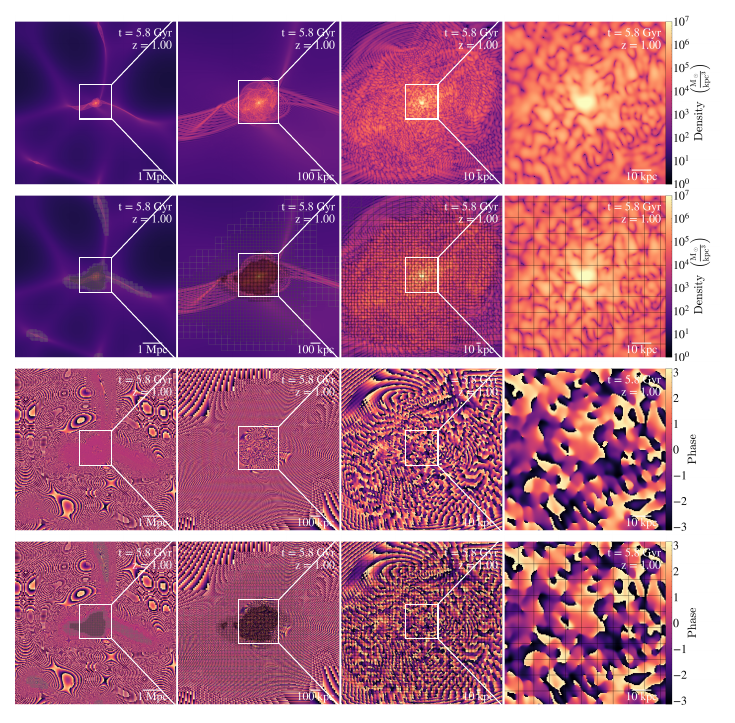}
\caption{Zoom into a halo at $z=1$ in the cosmological simulation shown in Fig. \ref{fig:madelung_schroedinger_mismatch}. The hybrid scheme accurately captures fully nonlinear FDM dynamics across length scales ranging from several Mpc down to sub-kpc. Top to bottom: Density slice; density slice with refinement grid highlighting wave levels; phase slice; phase slice with refinement grid highlighting wave levels. Each square in the refinement grid corresponds to $16^2$ cells ($16^3$ cells in three dimensions).}
\label{fig:halo}
\end{figure*}

Fig. \ref{fig:halo} zooms into a halo at $z=1$ and demonstrates that the hybrid scheme accurately captures fully nonlinear FDM dynamics across length scales ranging from several Mpc down to sub-kpc. The scheme also minimizes numerical artifacts through the wave and fluid algorithms and the interpolation scheme presented in this work. 

\begin{figure*}[ht!]
\centering
\includegraphics[width=\textwidth]{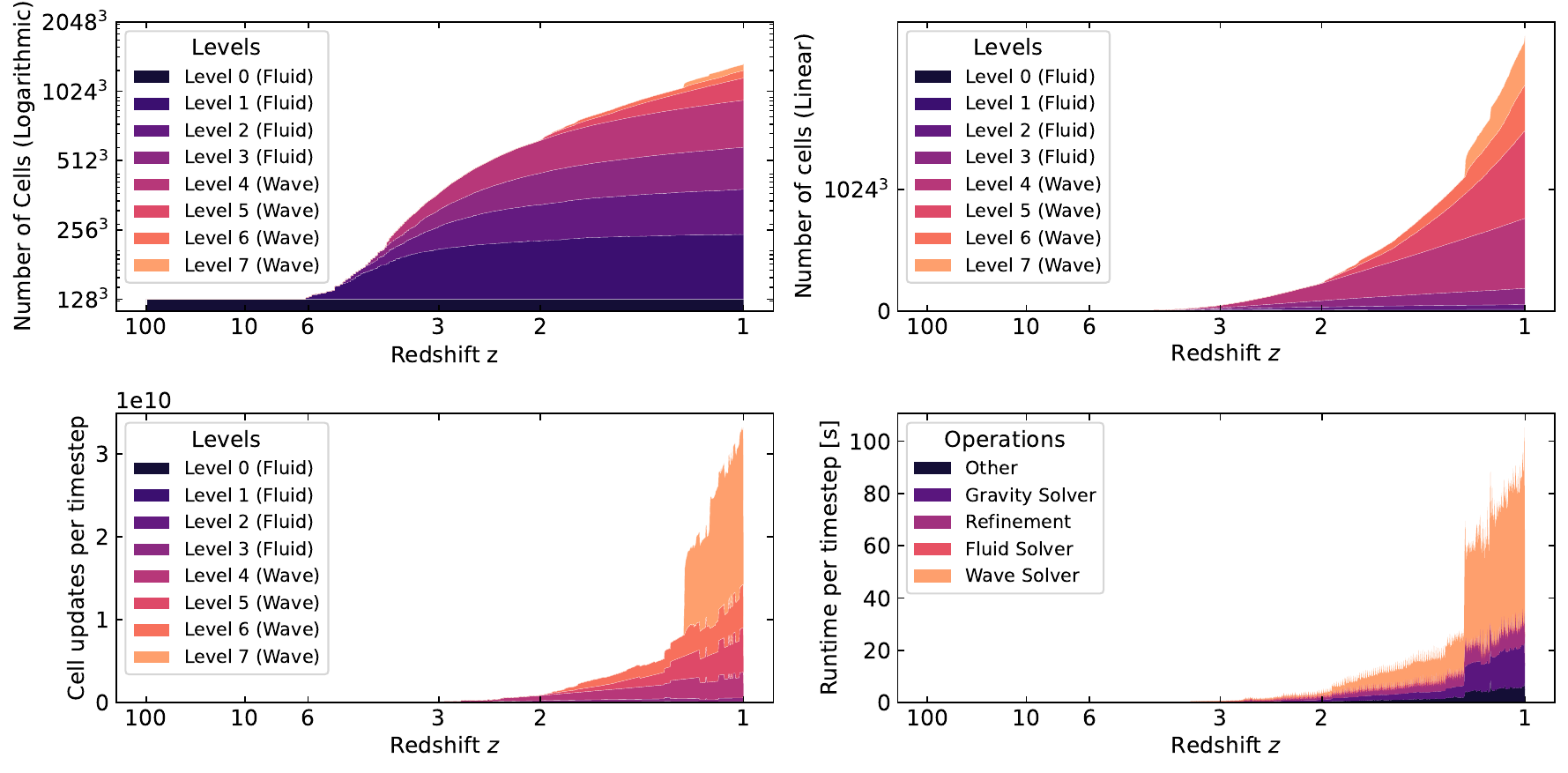}
\caption{Number of cells, cell updates, and runtime as a function of redshift for the cosmological simulation shown in Fig. \ref{fig:madelung_schroedinger_mismatch}. Most of the simulation time is consumed after $z=2$ for evolving patches using the wave solver on refinement levels $4$--$7$. The FC-Gram Matrix wave solver for the linear Schrödinger operation is the most time-consuming, followed by the refinement operation (flag patches for refinement + interpolation), the gravity solver, and other operations such as the time step calculation, MPI communication, and fix-up operations. In comparison, the fluid solver consumes a negligible amount of time. Note that data are presented cumulatively in this plot (e.g., in the upper-left panel, lines show contributions from level $0$, level $0$ + level $1$, and so on).}
\label{fig:runtime}
\end{figure*}

Fig. \ref{fig:runtime} shows the number of cells, the number of cell updates per root-level time step, and the runtime of various operations per root-level time step as a function of redshift $z$. At $z=1$, the majority of patches ($92\%$) use the wave solver, while $93\%$ of the simulation volume is evolved using the fluid solver. There are $4.8$ million patches in total, corresponding to $2.4\times 10^9 \sim 1346^3$ cells. Most of the simulation time is spent on evolving patches on wave levels $4-7$ using the FC-Gram Matrix algorithm. The relative mass conservation error is around $0.1\%$ at $z=1$. This serves as an important cross-check for the quality of the wave function evolution, refinement, and interpolation: Mass is not conserved on wave levels since the FC-Gram algorithm is not conservative, and additional mass loss occurs at the boundaries between patches of different resolutions. A sufficiently high resolution ensures that the FC-Gram algorithm does not dissipate mass beyond machine precision and the dynamics of neighboring patches with different resolutions agree sufficiently well.

\section{Discussion} \label{sec:discussion}
The hybrid scheme introduced in this work, combining fluid and wave solvers with AMR for the Schrödinger-Poisson equations, marks an advance over conventional approaches for simulating the dynamics of FDM. However, while the scheme offers noticeable improvements, it also presents several shortcomings that warrant further investigation. One of the persistent challenges in our current implementation is the presence of numerical artifacts. These issues arise primarily from spatial interpolation during grid refinement and when providing boundary conditions at the coarse-fine resolution interfaces, temporal interpolation for adaptive time steps, and insufficient refinement. We find that these artifacts can introduce unphysical small-scale waves, triggering unnecessary refinement. In addition, unphysical overdensities at high redshifts can source spurious halos, which are difficult to distinguish from genuine halos. Depending on the adopted simulation parameters, these spurious halos appear similar to the spurious halos in N-body simulations with initial power spectra involving suppression of high-$k$ modes, such as warm dark matter simulations \citep{Wang_2007} and collisionless N-body simulations with FDM initial conditions \citep{Schive2015}. This problem also appears in zoom-in simulations conducted with the hybrid scheme: Numerical artifacts at the boundary of the zoom-in region can contaminate the simulation with spurious halos if not treated properly. 

In terms of performance, our findings suggest that the performance of the FC-Gram Matrix algorithm is comparable to that of the finite difference method implemented in \texttt{GAMER}. The parallel scaling could potentially be further improved by adopting a larger patch size, which would reduce the amount of information exchange and communication overhead.

There is scope for improving the resolution power of the wave scheme. Switching to a Chebyshev method would improve the numerical conditioning of the algorithm and could further decrease the need for refinement. Our tests indicate that the Chebyshev method exhibits better convergence properties compared to the FC-Gram method. However, switching to a non-uniform mesh would require a substantial re-design of the \texttt{GAMER} code and necessitate smaller time steps because of the narrower grid spacing of Chebyshev grids. Given that even with the hybrid scheme presented in this work, large-volume FDM cosmological simulations for FDM masses significantly higher than $10^{-22}$ eV arguably remain prohibitively expensive, the question arises whether implicit methods leveraging non-uniform grids could be explored to circumvent the prohibitive time step conditions of the Schrödinger equation while retaining high resolution in halos and filaments at moderate computational cost.

\section{Conclusion} \label{sec:conclusion}
This study demonstrates the feasibility of hybrid AMR simulations for the Schrödinger-Poisson system, paving the way for more detailed and extensive simulations of the FDM model. It combines a fluid solver for the Hamilton-Jacobi-Madelung equations with a FC-Gram wave solver for the Schrödinger equation. We find the scheme to be stable, accurate, and computationally efficient. Suitable refinement criteria ensure that the fluid solver is only employed in regions without vortices and that the resolution is sufficiently high for the wave solver. These criteria are complemented by a FC-Gram interpolation scheme that adaptively interpolates the density and phase instead of the real and imaginary parts of the wave function in vortex-free regions. The resulting AMR algorithm proves to be versatile and broadly applicable to a variety of Schrödinger equation simulations. This includes the linear and nonlinear Schrödinger equations, as well as other forms of FDM models involving attractive or repulsive interactions, multi-field FDM, and mixed dark matter scenarios.

\section*{Acknowledgments}
We thank Tzihong Chiueh for many insightful discussions, Guan-Ming Su for his support in conducting and evaluating CDM and FDM simulations, and Chun-Yen Chen for his significant contributions to the development of \texttt{GAMER}. \revtext{We also thank the National Center for High-performance Computing (NCHC) for providing computational and storage resources.} This work made significant use of many open-source software packages, including \textsc{Python}, \textsc{IPython}, \textsc{NumPy} \citep{Numpy}, \textsc{SciPy} \citep{Scipy}, \textsc{Matplotlib} \citep{Matplotlib}, \textsc{YT} \citep{YT}, and \textsc{GSL} \citep{GSL}. These are products of collaborative efforts by many independent developers from numerous institutions around the world. Their commitment to open science has helped to make this work possible. This research is partially supported by the National Science and Technology Council (NSTC) of Taiwan under Grant No. NSTC 111-2628-M-002-005-MY4 and the NTU Academic Research-Career Development Project under Grant No. NTU-CDP-113L7729.

\bibliographystyle{aasjournal}
\bibliography{references.bib}{}

\end{document}